\documentclass[prl,reprint,notitlepage,groupedaddress,floatfix,nofootinbib,superscriptaddress]{revtex4-2}
\usepackage[final]{graphicx}
\usepackage{natbib}
\usepackage{hyperref}
\usepackage{amssymb}
\usepackage{amsmath}
\usepackage{mathrsfs}
\usepackage{xcolor}
\usepackage{bm}
 
\makeatletter

\begin{document}
\title{Attractive trion-polariton nonlinearity due to Coulomb scattering}

\author{Kok Wee Song}
\affiliation{Department of Physics and Astronomy, University of Exeter, Stocker Road, Exeter EX4 4QL, United Kingdom}

\author{Salvatore Chiavazzo}
\affiliation{Department of Physics and Astronomy, University of Exeter, Stocker Road, Exeter EX4 4QL, United Kingdom}

\author{Ivan A. Shelykh}
\affiliation{Science Institute, University of Iceland, Dunhagi 3, IS-107, Reykjavik, Iceland}
\affiliation{Department of Physics and Engineering, ITMO University, St. Petersburg 197101, Russia}

\author{Oleksandr Kyriienko}
\affiliation{Department of Physics and Astronomy, University of Exeter, Stocker Road, Exeter EX4 4QL, United Kingdom}

\date{\today}

\begin{abstract}
We theoretically investigate the nonlinearity of trion-polaritons in a two-dimensional material that arises from Coulomb interaction between quasiparticles. To evaluate the interaction constant, we solve a three-body Wannier equation precisely by expanding trion wavefunctions into a Gaussian basis. Using these wavefunctions, we calculate the trion-polariton interaction energies for the exchange processes, resolving the outstanding question of trion-trion scattering. We find that the nonlinearity is the result of the competition between different scattering channels. Such a cancellation effect is sensitive to wavefunction overlaps and depends on material parameters. Most importantly, our result shows that the nonlinear interaction between trion-polaritons is attractive, and is fivefold stronger than exciton-polariton interaction. Our work thus describes the regime where trion-polaritons offer the prospects for attractive fluids of light in monolayers of transition metal dichalcogenides. 
\end{abstract} 

\maketitle

% =====================
% Introduction
%======================

\textit{Introduction.---}Optical response of two-dimensional (2D) transition metal dichalcogenides (TMDs) is governed by strongly-bound composite particles~\cite{Mak:PRL105(2010),Splendiani:NanoLett10(2010),Chernikov:PRL115(2015),Ugeda:NatMat:13(2014),He:PRL113(2014),Zhang:PRB89(2014),You:NatPhys11(2015),Singh:PRB93(2016),Lundt:APL112(2018),Deilmann_2019}. The unique combination of large reduced mass \cite{Finland2015,Kormanyos2015,Mostaani2017}, valley physics \cite{Yu2015,Mueller2018}, and peculiar 2D screening \cite{Chernikov:PRL113(2014),Berkelbach:PRB88(2013)} leads to the dominant excitonic response, even at room temperature \cite{Wang:RMP90(2018)}. Large binding energy ($\sim300$~meV) and small size ($\sim1$~nm) of TMD excitons enables strong light-matter coupling (SC) to cavity photons \cite{Sidler:NatPhys13(2016),Dufferwiel:NatPhoto11(2017),Cuadra:NanoLett18(2018),Tan:PRX10(2020),Dhara:NPhys14(2017),Emmanuele:NatCommun11(2020),Kravtsov:LightSci9(2020),Zhang2018,Fernandez2019,Liu2015,Lackner2021} and plasmons \cite{Liu:NanoLett16(2016),Wen:NanoLett17(2017),Abid:ACSPhoto4(2017),Stuhrenberg2018,Yankovich2019,Goncalves2020}, leading to emergent exciton-polaritons \cite{Schneider2018,Huang_2022,CarusottoCiuti2013,BasovAsenjo2021}. Their hybrid nature enables nonlinear optical response coming from exciton-exciton (X-X) interaction \cite{Shahnazaryan:PRB96(2017),Barachati:NatNano13(2018),Bleu2020,SimPRB2020,Gu2021,Erkensten2021,Stepanov:PRL126(2021),Anton-Solanas2021}, though ultimately limited by small exciton size.

The optical response of \emph{doped} TMD monolayers reveals the presence of excitons strongly correlated with the Fermi sea, leading to a trion peak being routinely resolved in experiments \cite{Mak:NatMat12(2012),Ross:NatComm4(2013),Dufferwiel:NatPhoto11(2017),Lundt:APL112(2018),Lyons2019,Emmanuele:NatCommun11(2020),Zipfel2020,Zipfel2022}. Unlike GaAs structures where the trion response is limited by weak ($1$-$2$~meV) binding \cite{Finkelstein:PRL74(1995),Glasberg1999,Rapaport:PRL84(2000),Rapaport:PRB63(2001),Teran2005}, TMD samples offer the significant trion binding energy ($20$-$30$~meV) \cite{Ross:NatComm4(2013),Finland2015,Ganchev:PRL114(2015),Efimkin:PRB95(2017),Courtade:PRB96(2017),ZhumagulovPRB2020} and actively participate in SC~\cite{Sidler:NatPhys13(2016),Dufferwiel:NatPhoto11(2017),Dhara:NPhys14(2017),Cuadra:NanoLett18(2018),Barachati:NatNano13(2018),Tan:PRX10(2020),Emmanuele:NatCommun11(2020),Kravtsov:LightSci9(2020),Stepanov:PRL126(2021)}. The associated trion-polaritons (TP) or exciton-polaron-polaritons also attracted significant theoretical attention \cite{Shiau_2017,Ravets:PRL120(2018),Chang:PRB98(2018),Levinsen:PRR1(2019),Rana:PRB102(2020),Rana:PRL126(2021),Koksal:PRR3(2021),Kyriienko:PRL125(2020),Zhumagulov:NPJ2022,BastarracheaMagnani:PRL126(2021),LiBleuPRL2021,LiBleuPRB2021,EfimkinPRB2021,Tiene2022,Denning:PRB105(2022),Denning:Arxiv(2021)}, with a major effort devoted to understanding their optical response. Intuitively, trion-polaritons shall also serve as a platform for studying strongly nonlinear behavior, thanks to excessive charge and larger size of quasiparticles. However, the theoretical understanding of trion-trion (T-T) interaction remains an open theoretical challenge, posed by an intricate problem~\cite{Combescot:PRX7(2017)} of many-body correlations and absence of simple trial wavefunction~\cite{Combescot:EurPhysJB79(2011),Shiau:PRB86(2012)}. Solving this problem requires finding a way for finding an accurate ground state wavefunction of trions and evaluating numerous interparticle scattering channels  that account for their composite nature.

In this letter, we present a theoretical study of the nonlinear response of trion-polaritons. We develop an approach for trion ground state description in terms of a large number of Gaussian basis functions, which is detailed in an accompanying paper \cite{Song2022b}. This allows for an accurate evaluation of scattering amplitudes. Using the composite boson theory, we derive and numerically calculate the trion-trion potential based on direct and exchange processes (a total of 108 diagrams). We find the attractive nonlinearity for trions that is an order of magnitude larger than X-X interaction. This sheds light on recent experimental results~\cite{Emmanuele:NatCommun11(2020)} revealing strong trion-polariton nonlinearity, and contributes to future nonlinear and quantum polaritonic devices.
%%%
\begin{figure}
    \centering
    \includegraphics[width=3.3in]{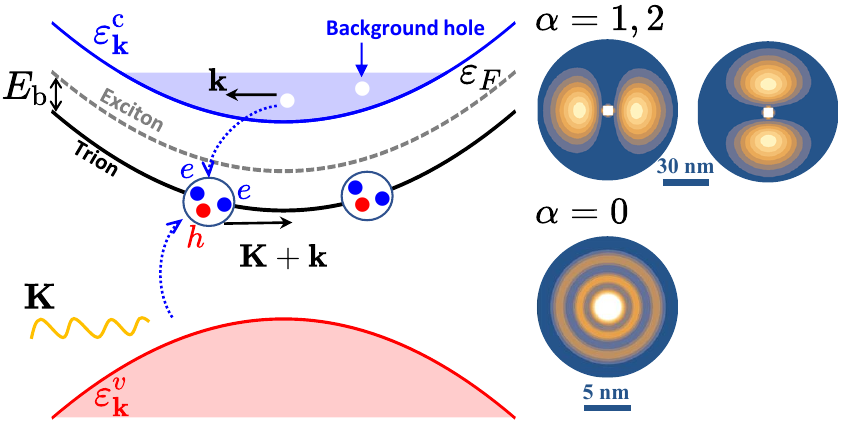}
    \caption{\textbf{(left)} Schematic diagram of a trion-polariton system. We show a partially-filled conduction band a valence band, where a photon creates electron-hole pairs that are correlated with electrons in the Fermi sea. %$U$ and $\tilde{U}$ are the effective trion-trion and trion-background hole interactions. 
    \textbf{(right)} Charge density plot for trion ground state with $E_b=31$~meV  ($\alpha=0$), and two lowest degenerate excited states ($\alpha=1,2$).}
    \label{fig:Trion}
\end{figure}
%%%

% =====================
% Trion in 2D materials
%======================
\textit{Model.---}We consider a model of interacting electrons and holes in a semiconductor monolayer described by two bands (Fig. \ref{fig:Trion}). The Hamiltonian reads
\begin{equation}\label{eqn:H}
    \mathcal{H}_e=\sum_{\mathbf{k}}(\varepsilon_\mathbf{k}^ca^\dagger_\mathbf{k}a_{\mathbf{k}}+\varepsilon_\mathbf{k}^vb^\dagger_\mathbf{k}b_{\mathbf{k}})+\frac{1}{2L^2}\sum_{\mathbf{q}}W(\mathbf{q})\rho_{\mathbf{q}}\rho_{-\mathbf{q}},
\end{equation}
where $L^2$ is an area of the system. Here $a^\dagger_{\mathbf{k}}$ and $b^\dagger_{\mathbf{k}}$ are creation field operators for the conduction and valence band electrons with momentum $\mathbf{k}$. $\rho_\mathbf{q}=\sum_{\mathbf{k}}(a^\dagger_{\mathbf{k}+\mathbf{q}}a_{\mathbf{k}}+b^\dagger_{\mathbf{k}+\mathbf{q}}b_{\mathbf{k}})$ is a total density operator. The interaction in 2D materials crucially takes account screening in the form of Keldysh-Rytova potential \cite{Keldysh:JETP29-1979,Rytova:MUPhys3-1967},
$
    W(\mathbf{q})=\frac{2\pi e^2/\epsilon}{q(1-r_\ast q)},
$
where $r_\ast$ is a screening length. We define $
    |0\rangle
$ as a ground state of $\mathcal{H}_e$ with no doping, and 
$
    |\varnothing\rangle=\prod_{\varepsilon^c(\mathbf{k})\leq \varepsilon_\mathrm{F}}a^\dagger_{\mathbf{k}}|0\rangle
$ as a ground state of doped system with Fermi energy $\varepsilon_\mathrm{F}$.
In our model, we consider the case of small doping, where photo-created excitons become correlated with electrons in the Fermi sea. The underlying quasiparticles are trion-polaritons that emerge from the hybridization between cavity photon and trion-based excitation described by a four-body operator~\cite{Rapaport:PRL84(2000), Rapaport:PRB63(2001)}
\begin{equation}
\label{eqn:B-Ta}
    \mathcal{B}^\dagger_{\alpha \mathbf{K}}=\frac{1}{\sqrt{N_\mathrm{F}}}\sum_{|\mathbf{k}|\leq k_\mathrm{F}}\mathcal{T}^\dagger_{\alpha,\mathbf{K}+\mathbf{k}}a_\mathbf{k},
\end{equation}
where $N_\mathrm{F}$ is the number of electrons in the Fermi sea and
\begin{equation}\label{eqn:T}
    \mathcal{T}^\dagger_{\alpha,\mathbf{K}+\mathbf{k}}=\sum_{\mathbf{k}_1\mathbf{k}_2\neq\mathbf{k}}\frac{1}{\sqrt{2!}}
    A^{\alpha,\mathbf{K}+\mathbf{k}}_{\mathbf{k}_1\mathbf{k}_2}a^\dagger_{\mathbf{k}_1}a^\dagger_{\mathbf{k}_2}b_{\mathbf{k}_1+\mathbf{k}_2-\mathbf{K}-\mathbf{k}}
\end{equation}
is the creation operator for a bound trion in a state $\alpha$. Here, we note that excitonic modes are explicitly excluded in the 4-body excitation by imposing momentum conservation for the sum with $\mathbf{k}_1,\mathbf{k}_2\neq \mathbf{k}$~\cite{Rana:PRB102(2020), Rana:PRL126(2021)}. In Eq. \eqref{eqn:B-Ta}, the excitation corresponds to creating a trion with momentum $\mathbf{K}+\mathbf{k}$ and leaving a hole in conduction band with momentum $-\mathbf{k}$ in the background (Fig.~\ref{fig:Trion}).
Finally, for samples placed in an optical microresonator, a cavity photon $c_{\mathbf{K}}$ can hybridize with the trion excitation $\mathcal{B}_{\alpha \mathbf{K}}$ through the linear coupling process, leading to trion-polaritons. 
The corresponding processes is described by the Hamiltonian~\cite{Kyriienko:PRL125(2020),Zhumagulov:NPJ2022,Rana:PRB102(2020),Rana:PRL126(2021)} 
\begin{equation}
\mathcal{H}_{\mathrm{SC}} = \sum_{\alpha \mathbf{K}} \frac{\Omega_\alpha}{2} c_{\mathbf{K}}^\dagger \mathcal{B}_{\alpha \mathbf{K}} + \mathrm{h.c},  
\end{equation}
where $\Omega_\alpha$ is a trion-polariton light-matter coupling constant that depends on a dipole moment of a transition, mode volume, trion wavefunction, and electron density. Then, a spectrum of the interacting trion-polariton system can be obtained by diagonalizing the full many-body Hamiltonian at SC. 

We note that trion-based excitations in Eq.~\eqref{eqn:B-Ta} have successfully described experimental results in III-V semiconductors~\cite{Rapaport:PRL84(2000), Rapaport:PRB63(2001)} and transition metal dichalcogenides~\cite{Emmanuele:NatCommun11(2020)}, where strong light-matter coupling shows superradiant scaling $\Omega \propto \sqrt{N_{\mathrm{F}}}$. This has been also confirmed with a full-scale numerical modelling~\cite{Zhumagulov:NPJ2022}, showing that both trion-polaritons and dressed exciton-polaritons emerge at non-zero doping. Similar physics can be described by Chevy ansatz~\cite{Chevy:PRA74(2006),Combescot:PRL101(2008),Efimkin:PRB103(2021)} leading to the exciton-polaron picture with attractive and repulsive branches~\cite{Sidler:NatPhys13(2016),Li:PRL126(2021),*Li:PRB103(2021)}, and analogous optical response. It was argued that the energy difference between these 4-body excitations from different approaches may be negligible in the $\varepsilon_\mathrm{F}\approx0$ limit~\cite{Glazov:JChemPhys153(2020)}. Moreover, recently a quasiparticle based on a 4-body bound state was considered, where the background hole and the trion are correlated in the lowest order~\cite{Rana:PRB102(2020),Rana:PRL126(2021),Koksal:PRR3(2021)}. 
However, as the bound state is shallow and does not change qualitatively the response, we consider Eq.~\eqref{eqn:B-Ta} a sensible choice for describing nonlinear effects in the system.

\textit{Trion-polariton wavefunction.---}To find the trion bound state in Eq.~\eqref{eqn:T}, we solve the Wannier equation for eigenevalues $E_{\alpha\mathbf{Q}}$, being~\cite{Combescot:PhysRep463(2008),Deilmann:PRL116(2016),Drueppel:NatComm8(2017),Torche:PRB100(2019),Zhumagulov:PRB101(2020),Zhumagulov:JChemPhys153(2020)}
\begin{align}\label{eqn:T-WannierQ}
(\varepsilon^{c}_{\mathbf{k}_1}\!+\!\varepsilon^{c}_{\mathbf{k}_2}\!-\!\varepsilon^{v}_{\mathbf{k}_1+\mathbf{k}_2-\mathbf{Q}})A^{\alpha \mathbf{Q}}_{\mathbf{k}_1\mathbf{k}_2}
+&\frac{1}{L^2}\sum_{{\mathbf{q}_1\mathbf{q}_2}}\Xi_{\mathbf{q}_1\mathbf{q}_2}A^{\alpha\mathbf{Q}}_{\mathbf{k}_1-\mathbf{q}_1,\mathbf{k}_2-\mathbf{q}_2}\notag\\
&
=
E_{\alpha\mathbf{Q}}A^{\alpha \mathbf{Q}}_{\mathbf{k}_1\mathbf{k}_2},
\end{align}
where the interacting kernel is 
$\Xi_{\mathbf{q}_1\mathbf{q}_2}=W(\mathbf{q}_2)\delta_{\mathbf{q}_1,-\mathbf{q}_2}-W(\mathbf{q}_1)\delta_{\mathbf{q}_2,0}-W(\mathbf{q}_2)\delta_{\mathbf{q}_1,0}$. We neglect the exchange interactions~\cite{Deilmann:PRL116(2016)} as their contribution is small. We diagonalize Eq. \eqref{eqn:T-WannierQ} by using the basis function method~\cite{Kidd:PRB93(2016)}, and expand $
    A^{\alpha\mathbf{Q}}_{\mathbf{k}_1\mathbf{k}_2}=\sum_{\mathbf{n}_1\mathbf{n}_2}C^{\alpha\mathbf{Q}}_{\mathbf{n}_1\mathbf{n}_2}\phi_{\mathbf{n}_1}(\mathbf{k}_1\lambda_1)\phi_{\mathbf{n}_2}(\mathbf{k}_2\lambda_2)
$ into a complete basis of harmonic oscillator functions~\cite{Ceferino:PRB101(2020)}. Here, $\phi_{\mathbf{n}}(\mathbf{k}\lambda)=\prod_{j=x,y}N_{n_j}\mathrm{e}^{-k_j^2\lambda^2/2}H_{n_j}(k_j\lambda)$ is an eigenstate of a 2D harmonic oscillator in $\mathbf{k}$-space, $H_{n_j}$ is a Hermite polynomial, and states are labelled by the vector $\mathbf{n}$ with components $n_j$. The normalization constant is $N_n=\sqrt{\lambda/(\pi^{1/2}2^nn!)}$, and $\lambda_{1,2}$ are adjustable parameters chosen by optimizing binding energies \cite{Song2022b}.

We solve Eq.~\eqref{eqn:T-WannierQ} for the case of WSe\textsubscript{2} monolayer, with parabolic bands $\varepsilon^c_{\mathbf{k}}=-\varepsilon^v_{\mathbf{k}}=\frac{k^2}{2m}$ with mass $m=0.32 m_0$ ($m_0$ is a free electron mass, $\hbar = 1$)~\cite{Courtade:PRB96(2017)}. The screening length is set to $r_\ast=4$~nm. We find that the lowest energy state with $\mathbf{Q}=0$ is $31$ meV lower than the excitonic ground state. %This result agrees to previous studies~\cite{Zhumagulov:NPJ2022}. 
In Fig.~\ref{fig:Trion} (right), we show the trion charge density \cite{Deilmann:PRL116(2016)} by using the real space wavefunction, $
\Phi^T_\mathbf{Q}(\mathbf{r}_1\mathbf{r}_2)= \sum_{\mathbf{k}_1\mathbf{k}_2}A_{\mathbf{k}_1\mathbf{k}_2}^{\alpha\mathbf{Q}}\frac{1}{2}[\mathrm{e}^{i(\mathbf{k}_1\cdot\mathbf{r}_1+\mathbf{k}_2\cdot\mathbf{r}_2)}-\mathrm{e}^{i(\mathbf{k}_1\cdot\mathbf{r}_2+\mathbf{k}_2\cdot\mathbf{r}_1)}]
$. The charge density is plotted as $\int d^2\mathbf{r}_1\int d^2\mathbf{r}_2\sum_{i=1,2}\delta(\mathbf{r}-\mathbf{r}_i)|\Phi(\mathbf{r}_1\mathbf{r}_2)|^2$. We note that in contrast to the exciton ground state, for trion excitations there is a charge-density modulation in the radial direction. The trion-polariton wavefunction emerges straightforwardly when considering a hybrid nature of the mode, with Hopfield coefficients coming from diagonalization of light-matter coupling Hamiltonian~\cite{CarusottoCiuti2013}.

\begin{figure}
    \centering
    \includegraphics[width=3.3in]{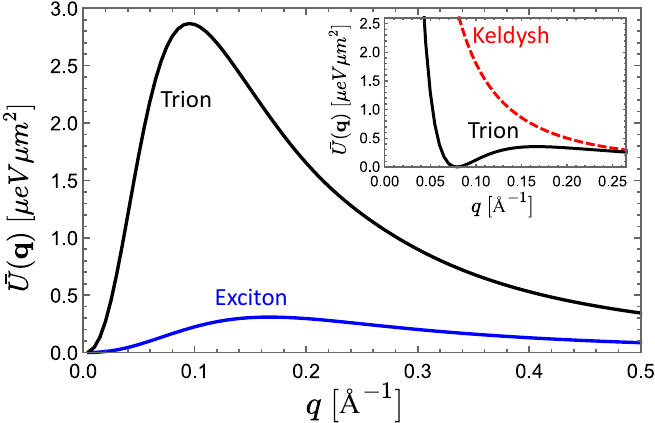}
    \caption{Comparison between direct interaction potentials. Black solid curve shows the direct trion-trion interaction $\bar{U}(\mathbf{q})$ with the subtracted background charge contribution. Blue solid curve shows the exciton-exciton effective interaction for the reference. \textbf{(inset)} Bare T-T interaction potential $U(\mathbf{q})$ without the subtraction of the background contributions (black solid), as well as the Keldysh potential of two interacting point-charges [red dashed, $W(\mathbf{q})$].}
    \label{fig:TTinteraction}
\end{figure}

% ========================
% trion-trion interactions
% ========================
\textit{Trion-trion interactions.---}We proceed to study the nonlinear response of trion-polaritons. Our goal is to describe the T-T interaction, which plays a crucial role in the nonlinearity of the TP spectrum. To estimate this effect, we derive the effective trion-trion interaction by using the approach based on creation potentials~\cite{Combescot:PhysRep463(2008),Combescot:EurPhysJB79(2011),Glazov2009}, which accounts implicitly for the composite nature of quasiparticles. First, we evaluate the commutator (see accompanying paper \cite{Song2022b} for the detailed derivation)
\begin{equation*}
    [\mathcal{H}_e,\mathcal{T}^\dagger_{\alpha\mathbf{Q}}]=E_{\alpha\mathbf{Q}}\mathcal{T}^\dagger_{\alpha\mathbf{Q}}\!+\!\frac{1}{L^2}\sum_{\alpha'\mathbf{q}}F^{\alpha\alpha'}_{\mathbf{Q},\mathbf{Q}+\mathbf{q}}W(\mathbf{q})\mathcal{T}^\dagger_{\alpha',\mathbf{Q}+\mathbf{q}}\rho_{-\mathbf{q}}
\end{equation*}
with $F^{\alpha\alpha'}_{\mathbf{Q},\mathbf{Q}+\mathbf{q}}\!=\!\displaystyle\sum_{\mathbf{k}_1\mathbf{k}_2}[A^{\alpha\mathbf{Q}}_{\mathbf{k}_1+\mathbf{q},\mathbf{k}_2}\!+\!A^{\alpha\mathbf{Q}}_{\mathbf{k}_1,\mathbf{k}_2+\mathbf{q}}
\!-\!A^{\alpha\mathbf{Q}}_{\mathbf{k}_1\mathbf{k}_2}]A^{\alpha',\mathbf{Q}+\mathbf{q},\ast}_{\mathbf{k}_1\mathbf{k}_2}$. The factor $F^{\alpha\alpha'}_{\mathbf{Q},\mathbf{Q}'}$ corresponds to the wavefunction overlap between the initial ($\alpha,\mathbf{Q}$) and final ($\alpha',\mathbf{Q}'$) states in the scattering process. This leads to
\begin{equation*}
    \{ [\mathcal{H}_e,\mathcal{T}^\dagger_{\alpha\mathbf{Q}}],\mathcal{T}^\dagger_{\beta\mathbf{P}}\}\!=\!\frac{1}{L^2}\!\!\sum_{\alpha'\beta'\mathbf{q}}\!U^{\mathbf{Q}\mathbf{P};\mathbf{Q}-\mathbf{q},\mathbf{P}+\mathbf{q}}_{\alpha\beta;\alpha'\beta'}\mathcal{T}^\dagger_{\alpha',\mathbf{Q}-\mathbf{q}}\mathcal{T}^\dagger_{\beta',\mathbf{P}+\mathbf{q}},
\end{equation*}
where the effective trion-trion interaction constant is
\begin{equation}\label{eqn:U}
    U^{\mathbf{Q}\mathbf{P};\mathbf{Q}'\mathbf{P}'}_{\alpha\beta;\alpha'\beta'}= F^{\alpha\alpha'}_{\mathbf{Q}\mathbf{Q}'}W(\mathbf{Q}-\mathbf{Q}')F^{\beta\beta'}_{\mathbf{P}\mathbf{P}'},
\end{equation}
and $(\mathbf{Q}-\mathbf{Q}')$ is the momentum transferred between trions. We note that Eq. \eqref{eqn:U} only accounts for the direct interaction between two ground state trions where no particles are exchanged, or the full trion-trion exchange happens (see Refs.~\cite{Ciuti1998,Glazov2009} for the analogous X-X processes). 
The corresponding dependence of $U(\mathbf{q})=U^{00;-\mathbf{q},\mathbf{q}}_{00;00}$ is plotted in the inset of Fig.~\ref{fig:TTinteraction}. We observe that the bare direct interaction potential has a long-range nature and in $q \rightarrow 0$ limit, we have $W(\mathbf{q})\sim1/q$, meaning that trions resemble point charges. However, as we show next the divergent contribution from direct T-T scattering is canceled by the background charge, and exchange interactions (those involving electron or hole swaps) are the key processes contributing to nonlinearity.
%%%
%%%

We demonstrate the effect of exchange interactions using an example of a two-trion state $|2\rangle=\mathcal{T}^\dagger_{\alpha\mathbf{Q}}\mathcal{T}^\dagger_{\beta\mathbf{P}}|0\rangle/R$ normalized by $R$. 
We derive the total energy of the system using anticommutation rules \cite{Combescot:PRL104(2010)}, $
    \{\mathcal{T}_{\alpha'\mathbf{Q}'},\mathcal{T}^\dagger_{\alpha\mathbf{Q}}\}=\delta_{\alpha'\alpha}\delta_{\mathbf{Q},\mathbf{Q}}+\mathcal{D}^{\alpha\alpha'}_{\mathbf{Q},\mathbf{Q}'}
$, where phase space filling operator $\mathcal{D}^{\alpha\alpha'}_{\mathbf{Q},\mathbf{Q}'}$ arises due to the composite nature of trions (see \cite{Song2022b} for the details). The total energy of this system is given by 
\begin{align}
    \langle2|\mathcal{H}_e|2\rangle=&E_{\alpha \mathbf{Q}}\!+\!E_{\beta \mathbf{P}}\!+[U_{\alpha\beta,\alpha\beta}^{\mathbf{Q}\mathbf{P},\mathbf{Q}\mathbf{P}}-U_{\alpha\beta,\beta\alpha}^{\mathbf{Q}\mathbf{P};\mathbf{P}\mathbf{Q}}\notag\\
    &+V^{\mathbf{Q}\mathbf{P};\mathbf{Q}\mathbf{P}}_{\alpha\beta;\alpha\beta}-V^{\mathbf{Q}\mathbf{P};\mathbf{P}\mathbf{Q}}_{\alpha\beta;\beta\alpha}]/L^2,
    \label{eqn:E2}
\end{align}
where the direct interactions $U^{\mathbf{Q}\mathbf{P};\mathbf{Q}'\mathbf{P}'}_{\alpha\beta;\alpha'\beta'}$ are the same as in Eq.~\eqref{eqn:U} and the exchange interactions are
\begin{equation}
    V^{\mathbf{Q}\mathbf{P},\mathbf{Q}'\mathbf{P}'}_{\alpha\beta,\alpha'\beta'}=\sum_{\mu\nu\mathbf{q}}U_{\alpha\beta,\mu\nu}^{\mathbf{Q}\mathbf{P};\mathbf{Q}-\mathbf{q},\mathbf{P}+\mathbf{q}}\Lambda^{\mu\nu,\alpha'\beta'}_{\mathbf{Q}-\mathbf{q},\mathbf{P}+\mathbf{q},\mathbf{Q}'\mathbf{P}'}
\end{equation}
with the Pauli scattering~\cite{Combescot:PhysRep463(2008),Glazov2009}
\begin{align}
    &\Lambda^{\alpha\beta,\alpha'\beta'}_{\mathbf{Q}\mathbf{P},\mathbf{Q}'\mathbf{P}'}\!=\!\sum_{\mathbf{k}_1\mathbf{k}_2}\sum_{\mathbf{p}_1\mathbf{p}_2}\Big(A^{\alpha'\mathbf{Q}'\ast}_{\mathbf{p}_1\mathbf{p}_2}\!A^{\beta'\mathbf{P}'\ast}_{\mathbf{k}_1\mathbf{k}_2}
    \delta_{\mathbf{p}_1+\mathbf{p}_2-\mathbf{Q}',\mathbf{k}_1+\mathbf{k}_2-\mathbf{Q}}
    \notag\\
    &
    -4A^{\alpha'\mathbf{Q}'\ast}_{\mathbf{k}_2\mathbf{p}_2}\!A^{\beta'\mathbf{P}'\ast}_{\mathbf{k}_1,\mathbf{p}_1}\delta_{\mathbf{p}_2-\mathbf{Q}',\mathbf{k}_1-\mathbf{Q}}
    \Big)A^{\alpha\mathbf{Q}}_{\mathbf{k}_1\mathbf{k}_2}\!A^{\beta\mathbf{P}}_{\mathbf{p}_1\mathbf{p}_2}.
    \label{eqn:Lambda}
\end{align}
This scattering function contains all the essential information about how two trion wavefunctions overlap in each different exchange process with $(\mathbf{Q}-\mathbf{Q}')$ transfer. 

\begin{figure}
    \centering
    \includegraphics[width=3.3in]{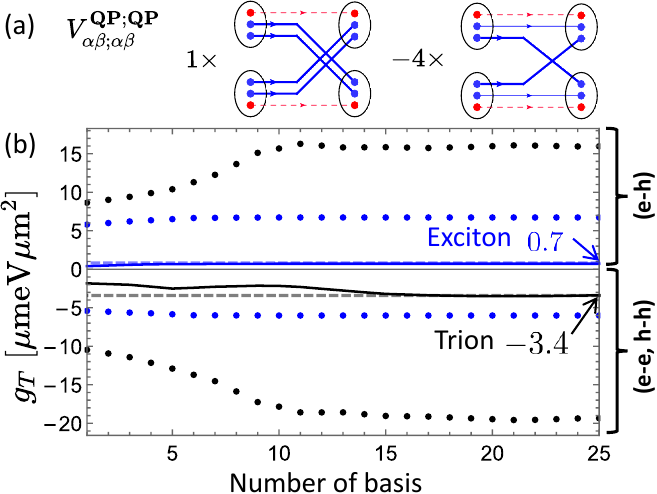}
    \caption{\textbf{(a)} Diagrammatic representation of exchange processes with zero momentum transfer, which represent the leading interaction processes. Coulomb interaction between electrons and holes in different trions leads to positive (e-h) and negative (e-e, h-h) exchange energy contributions. \textbf{(b)} Nonlinear interaction constant $g_\mathrm{T}=V^{00;00}_{00;00}$ shown as a function of basis set size, until convergence. Black points show separate contributions of T-T exchange based on e-h processes (top) and e-e/h-h processes (bottom). The total exchange interaction for two trions is shown by black solid curve, and is attractive, converging to $-3.4~\mu$eV$\mu$m$^2$. Blue points and blue solid curve show the same interaction for X-X scattering, converging to $0.7~\mu$eV$\mu$m$^2$ repulsive interaction.}
    \label{fig:Exchange}
\end{figure}

In Eq.~\eqref{eqn:E2}, the first line corresponds to trions treated as elementary fermions. The first two terms are one-trion energies. The third and fourth terms in \eqref{eqn:E2} correspond to direct T-T interaction (Hartree term) and full T-T exchange (Fock term)~\cite{Mahan:Many-Particle(2000)}, in analogy with X-X scattering \cite{Ciuti1998}. In the second line of \eqref{eqn:E2} we have energy corrections that arise purely from the exchange processes between constituents of two composite particles. They can be illustrated by the Shiva diagrams~\cite{Combescot:PhysRep463(2008)}, and we show example in Fig.~\ref{fig:Exchange}(a). In the diagram, we depict trions as black ellipsoids with electrons (blue dots) and holes (red dots) exchanging during the scattering process. The Coulomb interaction between electrons and holes of each composite particle can be divided into two groups, where e-h processes have a negative sign (attractive contribution) and e-e/h-h have a positive sign (repulsive contribution). 

Analysing the direct interaction, the bare Hartree term  $U^{\mathbf{Q}\mathbf{P};\mathbf{Q}\mathbf{P}}_{\alpha\beta;\alpha\beta}=W(0)$ formally diverges due to the long-range interaction between charged particles. However, this divergent behaviour is unphysical, as it does not account for the background charge (and since trions emerge from the Fermi sea same rules apply). We note that the divergence can be removed once the interaction with the background is taken into account. Intuitively, the renormalized interaction reads $\bar{U}(\mathbf{q})=U_{00,00}^{00;-\mathbf{q},\mathbf{q}}-2U_{\mathrm{T-h}}(\mathbf{q})+W(\mathbf{q})$, where $U_{\mathrm{T-h}}=F^{00}_{0,\mathbf{q}}W(\mathbf{q})$ is the trion-background hole interaction and $W(\mathbf{q})$ is the interaction between background charges (similar to the jellium model~\cite{Mahan:Many-Particle(2000),Song2022b}). We also find that cancellation appears naturally when $N$ trion-polaritons are considered together with Fermi sea. The dependence for trion-trion $\bar{U}(\mathbf{q})$ is plotted in Fig.~\eqref{fig:TTinteraction} as a black curve (main panel). We observe that at $q \approx 0$ the direct T-T interaction vanishes, similarly to the direct X-X case (see blue curve in Fig.~\eqref{fig:TTinteraction}). As nonlinear response of polaritons is defined by low-$q$ physics~\cite{CarusottoCiuti2013}, this does not contribute to the trion-polariton energy shift. The same applies to the renormalized Fock term, which has negligible contribution at $q \approx 0$ \cite{Ciuti1998,Song2022b}.
%%%
%%%

We proceed to calculate the exchange interaction. This involves 90 diagrams, separated into groups with zero and non-zero ($\mathbf{P}-\mathbf{Q}$) transferred momentum. As explained before, the first group makes the leading contribution [e.g. Fig.~\ref{fig:Exchange}(a)], and the second can be neglected. 
Therefore, we keep only the total energy correction with zero momentum transfer, $V^{\mathbf{Q}\mathbf{P};\mathbf{Q}\mathbf{P}}_{\alpha\beta;\alpha\beta}$ in Eq.~\eqref{eqn:E2}. Within this group of diagrams, we have interactions based on Coulomb attraction between opposite charges (e-h) and Coulomb repulsion for the same charges (e-e and h-h). We calculate numerically exchange sums~\cite{Song2022b}, showing these two subgroups for T-T scattering [Fig.~\ref{fig:Exchange}(b), black points], as well as X-X scattering [Fig.~\ref{fig:Exchange}(b), blue points] for the reference. We observe that e-h processes lead to the repulsive exchange energy contribution [Fig.~\ref{fig:Exchange}(b), top quadrant], while e-e/h-h processes lead to the attractive exchange contribution [Fig.~\ref{fig:Exchange}(b), bottom quadrant]. Combining the competing contributions, we observe that the T-T exchange is attractive overall (Fig.~\ref{fig:Exchange}(b), black curve), in strike difference to repulsive X-X exchange (Fig.~\ref{fig:Exchange}(b), blue curve)~\cite{Shahnazaryan:PRB96(2017)}. The possible origin of attraction may be tracked down to wavefunction with non-monotonic density (unlike 1s X), similarly to the case of excited X-X interaction \cite{Shahnazaryan:PRB102(2020),Shahnazaryan2016}. We note that the magnitude of interaction increases five-fold, and $g_\mathrm{T}$ is expected to grow further in systems with weaker binding.

% ==============
% N-Trion states
%===============
Next, we proceed to construct a state of $N$-trions that can efficiently hybridize with photons and form trion-polaritons \cite{Kyriienko:PRL125(2020)}. Using Eq.~\eqref{eqn:B-Ta}, this state can be written as 
$
    |N\rangle=(\mathcal{B}^\dagger_{\alpha \mathbf{K}})^N|\varnothing\rangle/\sqrt{N!F_N},
$
where the normalization factor is $N!F_N=\langle\varnothing|(\mathcal{B}_{\alpha \mathbf{K}})^N(\mathcal{B}^\dagger_{\alpha \mathbf{K}})^N|\varnothing\rangle$. This $N$-trion state has several desirable features. It ensures the cancellation of the divergence in the trion-trion direct interaction. Furthermore, it takes into account the conservation of momentum such that the created low-energy trion momentum is near to the photon momentum $\mathbf{K}$.

The total energy of $N$ trions is given by $E^{(N)}_{\text{tot}}=\langle N|\mathcal{H}_e|N\rangle$. We assume low electron density in the conduction band, and find that the total energy is
\begin{align}
    E^{(N)}_{\text{tot}}\approx&NE_{\alpha\mathbf{K}}+\frac{1}{L^2}\frac{F_{N-2}}{2F_N}\sum_{\mathbf{q}}W(\mathbf{q})\notag\\
    &\langle N-2|(\mathcal{B}_{\alpha\mathbf{K}})^2\mathcal{C}^\dagger_{\alpha\mathbf{K}}(-\mathbf{q})\mathcal{C}^\dagger_{\alpha\mathbf{K}}(\mathbf{q})|N-2\rangle ,
    \label{eqn:EN}
\end{align}
where $\mathcal{C}^\dagger_{\alpha\mathbf{K}}(\mathbf{q})=\frac{1}{\sqrt{N_\mathrm{F}}}\sum_{\alpha'\mathbf{k}}(F^{\alpha\alpha'}_{\mathbf{K}+\mathbf{k},\mathbf{K}+\mathbf{k}+\mathbf{q}}\mathcal{T}^\dagger_{\alpha',\mathbf{K}+\mathbf{k}+\mathbf{q}}a_{\mathbf{k}}-\delta_{\alpha\alpha'}\mathcal{T}^\dagger_{\alpha',\mathbf{K}+\mathbf{k}}a_{\mathbf{k}+\mathbf{q}})
$. The calculation of the expectation value is elaborate and is considered in the accompanying paper \cite{Song2022b}. The result is similar to the two-trion system shown in Eq. \eqref{eqn:E2}. However, we note that the $N$-trion case [Eq.~\eqref{eqn:EN}] contains three additional contributions. First is the interaction between trion and background hole that explicitly leads to the renormalized direct interaction, $\bar{U}(\mathbf{q})$. The second term includes exchange processes with more than two trion excitations~\cite{Combescot:PhysRep463(2008)}. The third term appears as an exchange interaction between trions and electrons in the Fermi sea, which leads to the renormalization of trions and background holes energies as a higher-order process~\cite{Efimkin:PRB95(2017),Glazov:JChemPhys153(2020)}. The second and third contributions are small in the low-density limit and remain beyond the scope of the current study. Thus, we keep only the pairwise T-T exchange processes and define
\begin{align}
    \Delta^{(N)}_{\mathbf{K}} =  E_{0,\mathbf{K}}+ng_\mathrm{T}+\mathcal{O}(n^2a_T^4),
    \label{eqn:nonlinearity}
\end{align}
as the lowest-order approximation for energy difference $\Delta^{(N)}_{\mathbf{K}}=E^{(N+1)}_{\text{tot}}-E^{(N)}_{\text{tot}}$ required for creating an additional trion in the $N$-trion system, and $n=N/L^2$ is a trion density. Here, we denote the nonlinearity as $g_\mathrm{T}=V^{00;00}_{00;00}$ being equal to exchange scattering amplitude of two ground state trions, and $a_T$ is a characteristic trion size. Eq.~\eqref{eqn:nonlinearity} describes the nonlinear redshift of optical response due to T-T scattering, which directly translates into trion-polariton nonlinearity. Namely, considering the polariton Hamiltonian for $N$ trions
\begin{equation}
    H=
    \begin{bmatrix}
        \Delta^{(N)}_{\mathbf{K}} & \Omega^{(N)}/2\\
        \Omega^{(N)}/2 & \omega_{c\mathbf{K}},
    \end{bmatrix}
\end{equation}
we can obtain the polariton eigenstates as a mixture of trion excitations with nonlinear energy spectrum and photons with equidistant spectrum with $\omega_{c\mathbf{K}}$ energy. The Rabi frequency $\Omega^{(N)}=2\langle N+1|\mathcal{H}_{\mathrm{SC}}c^\dagger_{\mathbf{K}}|N\rangle$ additionally depends on $N$ due to the phase space filling effect~\cite{Kyriienko:PRL125(2020)}. The resulting lower and upper polariton modes enjoy nonlinear shift due to both $\Delta^{(N)}$ and $\Omega^{(N)}$ dependence, being proportional to the trion density.

% ================
% discussion
%=================
\textit{Discussion.---}Recent experimental results revealed a strong nonlinear redshift when studying the upper polariton branch~\cite{Emmanuele:NatCommun11(2020)}, associated to Rabi splitting quenching. The major mechanism behind is the phase space filling for trion-polaritons---one has a limited number of TPs to be created, and the medium saturates in a nonlinear manner~\cite{Kyriienko:PRL125(2020),Yagafarov2020,Betzold2020}. At the same time, we highlight that for the TMD monolayer considered in our work the Coulomb-based trion-trion interaction is attractive, also leading to the redshift of polaritonic modes. Peculiarly, the two contributions for the upper TP add constructively, favoring the enhancement of the overall nonlinear response. Distinguishing the two requires an additional analysis of the nonlinear TP spectrum, as well as improved resolution for the lower polariton mode. We suggest this as a direction towards the systematic mapping of nonlinear contributions in doped TMD systems.

Next, the open question remains regarding the magnitude of the T-T scattering, and the universality of its sign. Are there materials and geometries where attraction changes into repulsion? Discovering systems where cancellation between scattering terms is strong (leading to repulsion) or weak (leading to large T-T interaction) will guide research on enhancing the nonlinearity \cite{Wild2018}.

Finally, as TMD exciton-polaritons enter the regime of condensation~\cite{Anton-Solanas2021}, this triggers the question of possible emergent coherence in trion-polariton systems. The X-X repulsion in III-V structures has led to dark solitons \cite{Sich2012,Walker2017,Maitre2020}, and vortices in quantum fluids of light~\cite{CarusottoCiuti2013}. Intriguingly, the attraction between trion-polaritons and the focusing nonlinearity may lead to attractive fluids of light hosted by gated semiconductor monolayers, resulting into polariton droplets \cite{Bradley1997,Almand-Hunter2014}, bright solitons~\cite{Khaykovich2002}, and photon molecules \cite{Firstenberg2013}. When brought to the quantum regime, this will offer exciting opportunities for quantum information processing \cite{KyriienkoLiew2016,Ghosh2021,XuLiew2021,Kuriakose2022}.

% ================
% Conclusion
%=================
\textit{Conclusion.---}We developed a theoretical approach for studying the Coulomb scattering processes between trion-polaritons, leading to the strongly nonlinear optical response. We employ a procedure for an accurate calculation of trion-polariton wavefunctions, decomposing them into the large-scale Gaussian basis, and evaluate interaction constants using the composite particle formalism. Considering a doped TMD monolayer, we find that the direct trion-trion scattering does not contribute to the nonlinear response due to cancellation with the background charge contribution. The exchange T-T scattering is enhanced up to an order of magnitude as compared to the X-X scattering, and has a negative sign. The attraction between trion-polaritons leads to nonlinear redshift, and may enable coherent effects for trion-polaritons.

\begin{acknowledgements}
\textit{Acknowledgements.---}The authors acknowledge the support from UK EPSRC New Investigator Award under the Agreement No. EP/V00171X/1. S.\,C. and O.\,K. are supported by the NATO Science for Peace and Security project NATO.SPS.MYP.G5860. I.\,A.\,S. acknowledges support from IRF (project ``Hybrid polaritonics'') and RFBR, project No. 21-52-12038.
\end{acknowledgements}

\bibliography{trion}

\begin{thebibliography}{117}%
\makeatletter
\providecommand \@ifxundefined [1]{%
 \@ifx{#1\undefined}
}%
\providecommand \@ifnum [1]{%
 \ifnum #1\expandafter \@firstoftwo
 \else \expandafter \@secondoftwo
 \fi
}%
\providecommand \@ifx [1]{%
 \ifx #1\expandafter \@firstoftwo
 \else \expandafter \@secondoftwo
 \fi
}%
\providecommand \natexlab [1]{#1}%
\providecommand \enquote  [1]{``#1''}%
\providecommand \bibnamefont  [1]{#1}%
\providecommand \bibfnamefont [1]{#1}%
\providecommand \citenamefont [1]{#1}%
\providecommand \href@noop [0]{\@secondoftwo}%
\providecommand \href [0]{\begingroup \@sanitize@url \@href}%
\providecommand \@href[1]{\@@startlink{#1}\@@href}%
\providecommand \@@href[1]{\endgroup#1\@@endlink}%
\providecommand \@sanitize@url [0]{\catcode `\\12\catcode `\$12\catcode
  `\&12\catcode `\#12\catcode `\^12\catcode `\_12\catcode `\%12\relax}%
\providecommand \@@startlink[1]{}%
\providecommand \@@endlink[0]{}%
\providecommand \url  [0]{\begingroup\@sanitize@url \@url }%
\providecommand \@url [1]{\endgroup\@href {#1}{\urlprefix }}%
\providecommand \urlprefix  [0]{URL }%
\providecommand \Eprint [0]{\href }%
\providecommand \doibase [0]{https://doi.org/}%
\providecommand \selectlanguage [0]{\@gobble}%
\providecommand \bibinfo  [0]{\@secondoftwo}%
\providecommand \bibfield  [0]{\@secondoftwo}%
\providecommand \translation [1]{[#1]}%
\providecommand \BibitemOpen [0]{}%
\providecommand \bibitemStop [0]{}%
\providecommand \bibitemNoStop [0]{.\EOS\space}%
\providecommand \EOS [0]{\spacefactor3000\relax}%
\providecommand \BibitemShut  [1]{\csname bibitem#1\endcsname}%
\let\auto@bib@innerbib\@empty
%</preamble>
\bibitem [{\citenamefont {Mak}\ \emph {et~al.}(2010)\citenamefont {Mak},
  \citenamefont {Lee}, \citenamefont {Hone}, \citenamefont {Shan},\ and\
  \citenamefont {Heinz}}]{Mak:PRL105(2010)}%
  \BibitemOpen
  \bibfield  {author} {\bibinfo {author} {\bibfnamefont {K.~F.}\ \bibnamefont
  {Mak}}, \bibinfo {author} {\bibfnamefont {C.}~\bibnamefont {Lee}}, \bibinfo
  {author} {\bibfnamefont {J.}~\bibnamefont {Hone}}, \bibinfo {author}
  {\bibfnamefont {J.}~\bibnamefont {Shan}},\ and\ \bibinfo {author}
  {\bibfnamefont {T.~F.}\ \bibnamefont {Heinz}},\ }\href
  {https://doi.org/10.1103/physrevlett.105.136805} {\bibfield  {journal}
  {\bibinfo  {journal} {Phys. Rev. Lett.}\ }\textbf {\bibinfo {volume} {105}},\
  \bibinfo {pages} {136805} (\bibinfo {year} {2010})}\BibitemShut {NoStop}%
\bibitem [{\citenamefont {Splendiani}\ \emph {et~al.}(2010)\citenamefont
  {Splendiani}, \citenamefont {Sun}, \citenamefont {Zhang}, \citenamefont {Li},
  \citenamefont {Kim}, \citenamefont {Chim}, \citenamefont {Galli},\ and\
  \citenamefont {Wang}}]{Splendiani:NanoLett10(2010)}%
  \BibitemOpen
  \bibfield  {author} {\bibinfo {author} {\bibfnamefont {A.}~\bibnamefont
  {Splendiani}}, \bibinfo {author} {\bibfnamefont {L.}~\bibnamefont {Sun}},
  \bibinfo {author} {\bibfnamefont {Y.}~\bibnamefont {Zhang}}, \bibinfo
  {author} {\bibfnamefont {T.}~\bibnamefont {Li}}, \bibinfo {author}
  {\bibfnamefont {J.}~\bibnamefont {Kim}}, \bibinfo {author} {\bibfnamefont
  {C.-Y.}\ \bibnamefont {Chim}}, \bibinfo {author} {\bibfnamefont
  {G.}~\bibnamefont {Galli}},\ and\ \bibinfo {author} {\bibfnamefont
  {F.}~\bibnamefont {Wang}},\ }\href {https://doi.org/10.1021/nl903868w}
  {\bibfield  {journal} {\bibinfo  {journal} {Nano Lett.}\ }\textbf {\bibinfo
  {volume} {10}},\ \bibinfo {pages} {1271} (\bibinfo {year}
  {2010})}\BibitemShut {NoStop}%
\bibitem [{\citenamefont {Chernikov}\ \emph {et~al.}(2015)\citenamefont
  {Chernikov}, \citenamefont {van~der Zande}, \citenamefont {Hill},
  \citenamefont {Rigosi}, \citenamefont {Velauthapillai}, \citenamefont
  {Hone},\ and\ \citenamefont {Heinz}}]{Chernikov:PRL115(2015)}%
  \BibitemOpen
  \bibfield  {author} {\bibinfo {author} {\bibfnamefont {A.}~\bibnamefont
  {Chernikov}}, \bibinfo {author} {\bibfnamefont {A.~M.}\ \bibnamefont {van~der
  Zande}}, \bibinfo {author} {\bibfnamefont {H.~M.}\ \bibnamefont {Hill}},
  \bibinfo {author} {\bibfnamefont {A.~F.}\ \bibnamefont {Rigosi}}, \bibinfo
  {author} {\bibfnamefont {A.}~\bibnamefont {Velauthapillai}}, \bibinfo
  {author} {\bibfnamefont {J.}~\bibnamefont {Hone}},\ and\ \bibinfo {author}
  {\bibfnamefont {T.~F.}\ \bibnamefont {Heinz}},\ }\href
  {https://doi.org/10.1103/physrevlett.115.126802} {\bibfield  {journal}
  {\bibinfo  {journal} {Phys. Rev. Lett.}\ }\textbf {\bibinfo {volume} {115}},\
  \bibinfo {pages} {126802} (\bibinfo {year} {2015})}\BibitemShut {NoStop}%
\bibitem [{\citenamefont {Ugeda}\ \emph {et~al.}(2014)\citenamefont {Ugeda},
  \citenamefont {Bradley}, \citenamefont {Shi}, \citenamefont {da~Jornada},
  \citenamefont {Zhang}, \citenamefont {Qiu}, \citenamefont {Ruan},
  \citenamefont {Mo}, \citenamefont {Hussain}, \citenamefont {Shen},
  \citenamefont {Wang}, \citenamefont {Louie},\ and\ \citenamefont
  {Crommie}}]{Ugeda:NatMat:13(2014)}%
  \BibitemOpen
  \bibfield  {author} {\bibinfo {author} {\bibfnamefont {M.~M.}\ \bibnamefont
  {Ugeda}}, \bibinfo {author} {\bibfnamefont {A.~J.}\ \bibnamefont {Bradley}},
  \bibinfo {author} {\bibfnamefont {S.-F.}\ \bibnamefont {Shi}}, \bibinfo
  {author} {\bibfnamefont {F.~H.}\ \bibnamefont {da~Jornada}}, \bibinfo
  {author} {\bibfnamefont {Y.}~\bibnamefont {Zhang}}, \bibinfo {author}
  {\bibfnamefont {D.~Y.}\ \bibnamefont {Qiu}}, \bibinfo {author} {\bibfnamefont
  {W.}~\bibnamefont {Ruan}}, \bibinfo {author} {\bibfnamefont {S.-K.}\
  \bibnamefont {Mo}}, \bibinfo {author} {\bibfnamefont {Z.}~\bibnamefont
  {Hussain}}, \bibinfo {author} {\bibfnamefont {Z.-X.}\ \bibnamefont {Shen}},
  \bibinfo {author} {\bibfnamefont {F.}~\bibnamefont {Wang}}, \bibinfo {author}
  {\bibfnamefont {S.~G.}\ \bibnamefont {Louie}},\ and\ \bibinfo {author}
  {\bibfnamefont {M.~F.}\ \bibnamefont {Crommie}},\ }\href
  {https://doi.org/10.1038/nmat4061} {\bibfield  {journal} {\bibinfo  {journal}
  {Nat. Mater.}\ }\textbf {\bibinfo {volume} {13}},\ \bibinfo {pages} {1091}
  (\bibinfo {year} {2014})}\BibitemShut {NoStop}%
\bibitem [{\citenamefont {He}\ \emph {et~al.}(2014)\citenamefont {He},
  \citenamefont {Kumar}, \citenamefont {Zhao}, \citenamefont {Wang},
  \citenamefont {Mak}, \citenamefont {Zhao},\ and\ \citenamefont
  {Shan}}]{He:PRL113(2014)}%
  \BibitemOpen
  \bibfield  {author} {\bibinfo {author} {\bibfnamefont {K.}~\bibnamefont
  {He}}, \bibinfo {author} {\bibfnamefont {N.}~\bibnamefont {Kumar}}, \bibinfo
  {author} {\bibfnamefont {L.}~\bibnamefont {Zhao}}, \bibinfo {author}
  {\bibfnamefont {Z.}~\bibnamefont {Wang}}, \bibinfo {author} {\bibfnamefont
  {K.~F.}\ \bibnamefont {Mak}}, \bibinfo {author} {\bibfnamefont
  {H.}~\bibnamefont {Zhao}},\ and\ \bibinfo {author} {\bibfnamefont
  {J.}~\bibnamefont {Shan}},\ }\href
  {https://doi.org/10.1103/physrevlett.113.026803} {\bibfield  {journal}
  {\bibinfo  {journal} {Phys. Rev. Lett.}\ }\textbf {\bibinfo {volume} {113}},\
  \bibinfo {pages} {026803} (\bibinfo {year} {2014})}\BibitemShut {NoStop}%
\bibitem [{\citenamefont {Zhang}\ \emph {et~al.}(2014)\citenamefont {Zhang},
  \citenamefont {Wang}, \citenamefont {Chan}, \citenamefont {Manolatou},\ and\
  \citenamefont {Rana}}]{Zhang:PRB89(2014)}%
  \BibitemOpen
  \bibfield  {author} {\bibinfo {author} {\bibfnamefont {C.}~\bibnamefont
  {Zhang}}, \bibinfo {author} {\bibfnamefont {H.}~\bibnamefont {Wang}},
  \bibinfo {author} {\bibfnamefont {W.}~\bibnamefont {Chan}}, \bibinfo {author}
  {\bibfnamefont {C.}~\bibnamefont {Manolatou}},\ and\ \bibinfo {author}
  {\bibfnamefont {F.}~\bibnamefont {Rana}},\ }\href
  {https://doi.org/10.1103/physrevb.89.205436} {\bibfield  {journal} {\bibinfo
  {journal} {Phys. Rev. B}\ }\textbf {\bibinfo {volume} {89}},\ \bibinfo
  {pages} {205436} (\bibinfo {year} {2014})}\BibitemShut {NoStop}%
\bibitem [{\citenamefont {You}\ \emph {et~al.}(2015)\citenamefont {You},
  \citenamefont {Zhang}, \citenamefont {Berkelbach}, \citenamefont {Hybertsen},
  \citenamefont {Reichman},\ and\ \citenamefont {Heinz}}]{You:NatPhys11(2015)}%
  \BibitemOpen
  \bibfield  {author} {\bibinfo {author} {\bibfnamefont {Y.}~\bibnamefont
  {You}}, \bibinfo {author} {\bibfnamefont {X.-X.}\ \bibnamefont {Zhang}},
  \bibinfo {author} {\bibfnamefont {T.~C.}\ \bibnamefont {Berkelbach}},
  \bibinfo {author} {\bibfnamefont {M.~S.}\ \bibnamefont {Hybertsen}}, \bibinfo
  {author} {\bibfnamefont {D.~R.}\ \bibnamefont {Reichman}},\ and\ \bibinfo
  {author} {\bibfnamefont {T.~F.}\ \bibnamefont {Heinz}},\ }\href
  {https://doi.org/10.1038/nphys3324} {\bibfield  {journal} {\bibinfo
  {journal} {Nat. Phys.}\ }\textbf {\bibinfo {volume} {11}},\ \bibinfo {pages}
  {477} (\bibinfo {year} {2015})}\BibitemShut {NoStop}%
\bibitem [{\citenamefont {Singh}\ \emph {et~al.}(2016)\citenamefont {Singh},
  \citenamefont {Moody}, \citenamefont {Tran}, \citenamefont {Scott},
  \citenamefont {Overbeck}, \citenamefont {Berghäuser}, \citenamefont
  {Schaibley}, \citenamefont {Seifert}, \citenamefont {Pleskot}, \citenamefont
  {Gabor}, \citenamefont {Yan}, \citenamefont {Mandrus}, \citenamefont
  {Richter}, \citenamefont {Malic}, \citenamefont {Xu},\ and\ \citenamefont
  {Li}}]{Singh:PRB93(2016)}%
  \BibitemOpen
  \bibfield  {author} {\bibinfo {author} {\bibfnamefont {A.}~\bibnamefont
  {Singh}}, \bibinfo {author} {\bibfnamefont {G.}~\bibnamefont {Moody}},
  \bibinfo {author} {\bibfnamefont {K.}~\bibnamefont {Tran}}, \bibinfo {author}
  {\bibfnamefont {M.~E.}\ \bibnamefont {Scott}}, \bibinfo {author}
  {\bibfnamefont {V.}~\bibnamefont {Overbeck}}, \bibinfo {author}
  {\bibfnamefont {G.}~\bibnamefont {Berghäuser}}, \bibinfo {author}
  {\bibfnamefont {J.}~\bibnamefont {Schaibley}}, \bibinfo {author}
  {\bibfnamefont {E.~J.}\ \bibnamefont {Seifert}}, \bibinfo {author}
  {\bibfnamefont {D.}~\bibnamefont {Pleskot}}, \bibinfo {author} {\bibfnamefont
  {N.~M.}\ \bibnamefont {Gabor}}, \bibinfo {author} {\bibfnamefont
  {J.}~\bibnamefont {Yan}}, \bibinfo {author} {\bibfnamefont {D.~G.}\
  \bibnamefont {Mandrus}}, \bibinfo {author} {\bibfnamefont {M.}~\bibnamefont
  {Richter}}, \bibinfo {author} {\bibfnamefont {E.}~\bibnamefont {Malic}},
  \bibinfo {author} {\bibfnamefont {X.}~\bibnamefont {Xu}},\ and\ \bibinfo
  {author} {\bibfnamefont {X.}~\bibnamefont {Li}},\ }\href
  {https://doi.org/10.1103/physrevb.93.041401} {\bibfield  {journal} {\bibinfo
  {journal} {Phys. Rev. B}\ }\textbf {\bibinfo {volume} {93}},\ \bibinfo
  {pages} {041401} (\bibinfo {year} {2016})}\BibitemShut {NoStop}%
\bibitem [{\citenamefont {Lundt}\ \emph {et~al.}(2018)\citenamefont {Lundt},
  \citenamefont {Cherotchenko}, \citenamefont {Iff}, \citenamefont {Fan},
  \citenamefont {Shen}, \citenamefont {Bigenwald}, \citenamefont {Kavokin},
  \citenamefont {Höfling},\ and\ \citenamefont
  {Schneider}}]{Lundt:APL112(2018)}%
  \BibitemOpen
  \bibfield  {author} {\bibinfo {author} {\bibfnamefont {N.}~\bibnamefont
  {Lundt}}, \bibinfo {author} {\bibfnamefont {E.}~\bibnamefont {Cherotchenko}},
  \bibinfo {author} {\bibfnamefont {O.}~\bibnamefont {Iff}}, \bibinfo {author}
  {\bibfnamefont {X.}~\bibnamefont {Fan}}, \bibinfo {author} {\bibfnamefont
  {Y.}~\bibnamefont {Shen}}, \bibinfo {author} {\bibfnamefont {P.}~\bibnamefont
  {Bigenwald}}, \bibinfo {author} {\bibfnamefont {A.~V.}\ \bibnamefont
  {Kavokin}}, \bibinfo {author} {\bibfnamefont {S.}~\bibnamefont {Höfling}},\
  and\ \bibinfo {author} {\bibfnamefont {C.}~\bibnamefont {Schneider}},\ }\href
  {https://doi.org/10.1063/1.5019177} {\bibfield  {journal} {\bibinfo
  {journal} {Appl. Phys. Lett.}\ }\textbf {\bibinfo {volume} {112}},\ \bibinfo
  {pages} {031107} (\bibinfo {year} {2018})}\BibitemShut {NoStop}%
\bibitem [{\citenamefont {Deilmann}\ and\ \citenamefont
  {Thygesen}(2019)}]{Deilmann_2019}%
  \BibitemOpen
  \bibfield  {author} {\bibinfo {author} {\bibfnamefont {T.}~\bibnamefont
  {Deilmann}}\ and\ \bibinfo {author} {\bibfnamefont {K.~S.}\ \bibnamefont
  {Thygesen}},\ }\href {https://doi.org/10.1088/2053-1583/ab0e1d} {\bibfield
  {journal} {\bibinfo  {journal} {2D Materials}\ }\textbf {\bibinfo {volume}
  {6}},\ \bibinfo {pages} {035003} (\bibinfo {year} {2019})}\BibitemShut
  {NoStop}%
\bibitem [{\citenamefont {Kyl\"anp\"a\"a}\ and\ \citenamefont
  {Komsa}(2015)}]{Finland2015}%
  \BibitemOpen
  \bibfield  {author} {\bibinfo {author} {\bibfnamefont {I.}~\bibnamefont
  {Kyl\"anp\"a\"a}}\ and\ \bibinfo {author} {\bibfnamefont {H.-P.}\
  \bibnamefont {Komsa}},\ }\href {https://doi.org/10.1103/PhysRevB.92.205418}
  {\bibfield  {journal} {\bibinfo  {journal} {Phys. Rev. B}\ }\textbf {\bibinfo
  {volume} {92}},\ \bibinfo {pages} {205418} (\bibinfo {year}
  {2015})}\BibitemShut {NoStop}%
\bibitem [{\citenamefont {Korm{\'{a}}nyos}\ \emph {et~al.}(2015)\citenamefont
  {Korm{\'{a}}nyos}, \citenamefont {Burkard}, \citenamefont {Gmitra},
  \citenamefont {Fabian}, \citenamefont {Z{\'{o}}lyomi}, \citenamefont
  {Drummond},\ and\ \citenamefont {Fal'ko}}]{Kormanyos2015}%
  \BibitemOpen
  \bibfield  {author} {\bibinfo {author} {\bibfnamefont {A.}~\bibnamefont
  {Korm{\'{a}}nyos}}, \bibinfo {author} {\bibfnamefont {G.}~\bibnamefont
  {Burkard}}, \bibinfo {author} {\bibfnamefont {M.}~\bibnamefont {Gmitra}},
  \bibinfo {author} {\bibfnamefont {J.}~\bibnamefont {Fabian}}, \bibinfo
  {author} {\bibfnamefont {V.}~\bibnamefont {Z{\'{o}}lyomi}}, \bibinfo {author}
  {\bibfnamefont {N.~D.}\ \bibnamefont {Drummond}},\ and\ \bibinfo {author}
  {\bibfnamefont {V.}~\bibnamefont {Fal'ko}},\ }\href
  {https://doi.org/10.1088/2053-1583/2/2/022001} {\bibfield  {journal}
  {\bibinfo  {journal} {2D Materials}\ }\textbf {\bibinfo {volume} {2}},\
  \bibinfo {pages} {022001} (\bibinfo {year} {2015})}\BibitemShut {NoStop}%
\bibitem [{\citenamefont {Mostaani}\ \emph {et~al.}(2017)\citenamefont
  {Mostaani}, \citenamefont {Szyniszewski}, \citenamefont {Price},
  \citenamefont {Maezono}, \citenamefont {Danovich}, \citenamefont {Hunt},
  \citenamefont {Drummond},\ and\ \citenamefont {Fal'ko}}]{Mostaani2017}%
  \BibitemOpen
  \bibfield  {author} {\bibinfo {author} {\bibfnamefont {E.}~\bibnamefont
  {Mostaani}}, \bibinfo {author} {\bibfnamefont {M.}~\bibnamefont
  {Szyniszewski}}, \bibinfo {author} {\bibfnamefont {C.~H.}\ \bibnamefont
  {Price}}, \bibinfo {author} {\bibfnamefont {R.}~\bibnamefont {Maezono}},
  \bibinfo {author} {\bibfnamefont {M.}~\bibnamefont {Danovich}}, \bibinfo
  {author} {\bibfnamefont {R.~J.}\ \bibnamefont {Hunt}}, \bibinfo {author}
  {\bibfnamefont {N.~D.}\ \bibnamefont {Drummond}},\ and\ \bibinfo {author}
  {\bibfnamefont {V.~I.}\ \bibnamefont {Fal'ko}},\ }\href
  {https://doi.org/10.1103/PhysRevB.96.075431} {\bibfield  {journal} {\bibinfo
  {journal} {Phys. Rev. B}\ }\textbf {\bibinfo {volume} {96}},\ \bibinfo
  {pages} {075431} (\bibinfo {year} {2017})}\BibitemShut {NoStop}%
\bibitem [{\citenamefont {Yu}\ \emph {et~al.}(2015)\citenamefont {Yu},
  \citenamefont {Cui}, \citenamefont {Xu},\ and\ \citenamefont {Yao}}]{Yu2015}%
  \BibitemOpen
  \bibfield  {author} {\bibinfo {author} {\bibfnamefont {H.}~\bibnamefont
  {Yu}}, \bibinfo {author} {\bibfnamefont {X.}~\bibnamefont {Cui}}, \bibinfo
  {author} {\bibfnamefont {X.}~\bibnamefont {Xu}},\ and\ \bibinfo {author}
  {\bibfnamefont {W.}~\bibnamefont {Yao}},\ }\href
  {https://doi.org/10.1093/nsr/nwu078} {\bibfield  {journal} {\bibinfo
  {journal} {Natl. Sci. Rev.}\ }\textbf {\bibinfo {volume} {2}},\ \bibinfo
  {pages} {57} (\bibinfo {year} {2015})}\BibitemShut {NoStop}%
\bibitem [{\citenamefont {Mueller}\ and\ \citenamefont
  {Malic}(2018)}]{Mueller2018}%
  \BibitemOpen
  \bibfield  {author} {\bibinfo {author} {\bibfnamefont {T.}~\bibnamefont
  {Mueller}}\ and\ \bibinfo {author} {\bibfnamefont {E.}~\bibnamefont
  {Malic}},\ }\href {https://doi.org/10.1038/s41699-018-0074-2} {\bibfield
  {journal} {\bibinfo  {journal} {npj 2D Materials and Applications}\ }\textbf
  {\bibinfo {volume} {2}},\ \bibinfo {pages} {29} (\bibinfo {year}
  {2018})}\BibitemShut {NoStop}%
\bibitem [{\citenamefont {Chernikov}\ \emph {et~al.}(2014)\citenamefont
  {Chernikov}, \citenamefont {Berkelbach}, \citenamefont {Hill}, \citenamefont
  {Rigosi}, \citenamefont {Li}, \citenamefont {Aslan}, \citenamefont
  {Reichman}, \citenamefont {Hybertsen},\ and\ \citenamefont
  {Heinz}}]{Chernikov:PRL113(2014)}%
  \BibitemOpen
  \bibfield  {author} {\bibinfo {author} {\bibfnamefont {A.}~\bibnamefont
  {Chernikov}}, \bibinfo {author} {\bibfnamefont {T.~C.}\ \bibnamefont
  {Berkelbach}}, \bibinfo {author} {\bibfnamefont {H.~M.}\ \bibnamefont
  {Hill}}, \bibinfo {author} {\bibfnamefont {A.}~\bibnamefont {Rigosi}},
  \bibinfo {author} {\bibfnamefont {Y.}~\bibnamefont {Li}}, \bibinfo {author}
  {\bibfnamefont {O.~B.}\ \bibnamefont {Aslan}}, \bibinfo {author}
  {\bibfnamefont {D.~R.}\ \bibnamefont {Reichman}}, \bibinfo {author}
  {\bibfnamefont {M.~S.}\ \bibnamefont {Hybertsen}},\ and\ \bibinfo {author}
  {\bibfnamefont {T.~F.}\ \bibnamefont {Heinz}},\ }\href
  {https://doi.org/10.1103/physrevlett.113.076802} {\bibfield  {journal}
  {\bibinfo  {journal} {Phys. Rev. Lett.}\ }\textbf {\bibinfo {volume} {113}},\
  \bibinfo {pages} {076802} (\bibinfo {year} {2014})}\BibitemShut {NoStop}%
\bibitem [{\citenamefont {Berkelbach}\ \emph {et~al.}(2013)\citenamefont
  {Berkelbach}, \citenamefont {Hybertsen},\ and\ \citenamefont
  {Reichman}}]{Berkelbach:PRB88(2013)}%
  \BibitemOpen
  \bibfield  {author} {\bibinfo {author} {\bibfnamefont {T.~C.}\ \bibnamefont
  {Berkelbach}}, \bibinfo {author} {\bibfnamefont {M.~S.}\ \bibnamefont
  {Hybertsen}},\ and\ \bibinfo {author} {\bibfnamefont {D.~R.}\ \bibnamefont
  {Reichman}},\ }\href {https://doi.org/10.1103/physrevb.88.045318} {\bibfield
  {journal} {\bibinfo  {journal} {Phys. Rev. B}\ }\textbf {\bibinfo {volume}
  {88}},\ \bibinfo {pages} {045318} (\bibinfo {year} {2013})}\BibitemShut
  {NoStop}%
\bibitem [{\citenamefont {Wang}\ \emph {et~al.}(2018)\citenamefont {Wang},
  \citenamefont {Chernikov}, \citenamefont {Glazov}, \citenamefont {Heinz},
  \citenamefont {Marie}, \citenamefont {Amand},\ and\ \citenamefont
  {Urbaszek}}]{Wang:RMP90(2018)}%
  \BibitemOpen
  \bibfield  {author} {\bibinfo {author} {\bibfnamefont {G.}~\bibnamefont
  {Wang}}, \bibinfo {author} {\bibfnamefont {A.}~\bibnamefont {Chernikov}},
  \bibinfo {author} {\bibfnamefont {M.~M.}\ \bibnamefont {Glazov}}, \bibinfo
  {author} {\bibfnamefont {T.~F.}\ \bibnamefont {Heinz}}, \bibinfo {author}
  {\bibfnamefont {X.}~\bibnamefont {Marie}}, \bibinfo {author} {\bibfnamefont
  {T.}~\bibnamefont {Amand}},\ and\ \bibinfo {author} {\bibfnamefont
  {B.}~\bibnamefont {Urbaszek}},\ }\href
  {https://doi.org/10.1103/revmodphys.90.021001} {\bibfield  {journal}
  {\bibinfo  {journal} {Reviews of Modern Physics}\ }\textbf {\bibinfo {volume}
  {90}},\ \bibinfo {pages} {021001} (\bibinfo {year} {2018})}\BibitemShut
  {NoStop}%
\bibitem [{\citenamefont {Sidler}\ \emph {et~al.}(2016)\citenamefont {Sidler},
  \citenamefont {Back}, \citenamefont {Cotlet}, \citenamefont {Srivastava},
  \citenamefont {Fink}, \citenamefont {Kroner}, \citenamefont {Demler},\ and\
  \citenamefont {Imamoglu}}]{Sidler:NatPhys13(2016)}%
  \BibitemOpen
  \bibfield  {author} {\bibinfo {author} {\bibfnamefont {M.}~\bibnamefont
  {Sidler}}, \bibinfo {author} {\bibfnamefont {P.}~\bibnamefont {Back}},
  \bibinfo {author} {\bibfnamefont {O.}~\bibnamefont {Cotlet}}, \bibinfo
  {author} {\bibfnamefont {A.}~\bibnamefont {Srivastava}}, \bibinfo {author}
  {\bibfnamefont {T.}~\bibnamefont {Fink}}, \bibinfo {author} {\bibfnamefont
  {M.}~\bibnamefont {Kroner}}, \bibinfo {author} {\bibfnamefont
  {E.}~\bibnamefont {Demler}},\ and\ \bibinfo {author} {\bibfnamefont
  {A.}~\bibnamefont {Imamoglu}},\ }\href {https://doi.org/10.1038/nphys3949}
  {\bibfield  {journal} {\bibinfo  {journal} {Nat. Phys.}\ }\textbf {\bibinfo
  {volume} {13}},\ \bibinfo {pages} {255} (\bibinfo {year} {2016})}\BibitemShut
  {NoStop}%
\bibitem [{\citenamefont {Dufferwiel}\ \emph {et~al.}(2017)\citenamefont
  {Dufferwiel}, \citenamefont {Lyons}, \citenamefont {Solnyshkov},
  \citenamefont {Trichet}, \citenamefont {Withers}, \citenamefont {Schwarz},
  \citenamefont {Malpuech}, \citenamefont {Smith}, \citenamefont {Novoselov},
  \citenamefont {Skolnick}, \citenamefont {Krizhanovskii},\ and\ \citenamefont
  {Tartakovskii}}]{Dufferwiel:NatPhoto11(2017)}%
  \BibitemOpen
  \bibfield  {author} {\bibinfo {author} {\bibfnamefont {S.}~\bibnamefont
  {Dufferwiel}}, \bibinfo {author} {\bibfnamefont {T.~P.}\ \bibnamefont
  {Lyons}}, \bibinfo {author} {\bibfnamefont {D.~D.}\ \bibnamefont
  {Solnyshkov}}, \bibinfo {author} {\bibfnamefont {A.~A.~P.}\ \bibnamefont
  {Trichet}}, \bibinfo {author} {\bibfnamefont {F.}~\bibnamefont {Withers}},
  \bibinfo {author} {\bibfnamefont {S.}~\bibnamefont {Schwarz}}, \bibinfo
  {author} {\bibfnamefont {G.}~\bibnamefont {Malpuech}}, \bibinfo {author}
  {\bibfnamefont {J.~M.}\ \bibnamefont {Smith}}, \bibinfo {author}
  {\bibfnamefont {K.~S.}\ \bibnamefont {Novoselov}}, \bibinfo {author}
  {\bibfnamefont {M.~S.}\ \bibnamefont {Skolnick}}, \bibinfo {author}
  {\bibfnamefont {D.~N.}\ \bibnamefont {Krizhanovskii}},\ and\ \bibinfo
  {author} {\bibfnamefont {A.~I.}\ \bibnamefont {Tartakovskii}},\ }\href
  {https://doi.org/10.1038/nphoton.2017.125} {\bibfield  {journal} {\bibinfo
  {journal} {Nat. Photon.}\ }\textbf {\bibinfo {volume} {11}},\ \bibinfo
  {pages} {497} (\bibinfo {year} {2017})}\BibitemShut {NoStop}%
\bibitem [{\citenamefont {Cuadra}\ \emph {et~al.}(2018)\citenamefont {Cuadra},
  \citenamefont {Baranov}, \citenamefont {Wersäll}, \citenamefont {Verre},
  \citenamefont {Antosiewicz},\ and\ \citenamefont
  {Shegai}}]{Cuadra:NanoLett18(2018)}%
  \BibitemOpen
  \bibfield  {author} {\bibinfo {author} {\bibfnamefont {J.}~\bibnamefont
  {Cuadra}}, \bibinfo {author} {\bibfnamefont {D.~G.}\ \bibnamefont {Baranov}},
  \bibinfo {author} {\bibfnamefont {M.}~\bibnamefont {Wersäll}}, \bibinfo
  {author} {\bibfnamefont {R.}~\bibnamefont {Verre}}, \bibinfo {author}
  {\bibfnamefont {T.~J.}\ \bibnamefont {Antosiewicz}},\ and\ \bibinfo {author}
  {\bibfnamefont {T.}~\bibnamefont {Shegai}},\ }\href
  {https://doi.org/10.1021/acs.nanolett.7b04965} {\bibfield  {journal}
  {\bibinfo  {journal} {Nano Lett.}\ }\textbf {\bibinfo {volume} {18}},\
  \bibinfo {pages} {1777} (\bibinfo {year} {2018})}\BibitemShut {NoStop}%
\bibitem [{\citenamefont {Tan}\ \emph {et~al.}(2020)\citenamefont {Tan},
  \citenamefont {Cotlet}, \citenamefont {Bergschneider}, \citenamefont
  {Schmidt}, \citenamefont {Back}, \citenamefont {Shimazaki}, \citenamefont
  {Kroner},\ and\ \citenamefont {{\.{I}}mamo{\u{g}}lu}}]{Tan:PRX10(2020)}%
  \BibitemOpen
  \bibfield  {author} {\bibinfo {author} {\bibfnamefont {L.~B.}\ \bibnamefont
  {Tan}}, \bibinfo {author} {\bibfnamefont {O.}~\bibnamefont {Cotlet}},
  \bibinfo {author} {\bibfnamefont {A.}~\bibnamefont {Bergschneider}}, \bibinfo
  {author} {\bibfnamefont {R.}~\bibnamefont {Schmidt}}, \bibinfo {author}
  {\bibfnamefont {P.}~\bibnamefont {Back}}, \bibinfo {author} {\bibfnamefont
  {Y.}~\bibnamefont {Shimazaki}}, \bibinfo {author} {\bibfnamefont
  {M.}~\bibnamefont {Kroner}},\ and\ \bibinfo {author} {\bibfnamefont
  {A.}~\bibnamefont {{\.{I}}mamo{\u{g}}lu}},\ }\href
  {https://doi.org/10.1103/physrevx.10.021011} {\bibfield  {journal} {\bibinfo
  {journal} {Phys. Rev. X}\ }\textbf {\bibinfo {volume} {10}},\ \bibinfo
  {pages} {021011} (\bibinfo {year} {2020})}\BibitemShut {NoStop}%
\bibitem [{\citenamefont {Dhara}\ \emph {et~al.}(2017)\citenamefont {Dhara},
  \citenamefont {Chakraborty}, \citenamefont {Goodfellow}, \citenamefont {Qiu},
  \citenamefont {O'Loughlin}, \citenamefont {Wicks}, \citenamefont
  {Bhattacharjee},\ and\ \citenamefont {Vamivakas}}]{Dhara:NPhys14(2017)}%
  \BibitemOpen
  \bibfield  {author} {\bibinfo {author} {\bibfnamefont {S.}~\bibnamefont
  {Dhara}}, \bibinfo {author} {\bibfnamefont {C.}~\bibnamefont {Chakraborty}},
  \bibinfo {author} {\bibfnamefont {K.~M.}\ \bibnamefont {Goodfellow}},
  \bibinfo {author} {\bibfnamefont {L.}~\bibnamefont {Qiu}}, \bibinfo {author}
  {\bibfnamefont {T.~A.}\ \bibnamefont {O'Loughlin}}, \bibinfo {author}
  {\bibfnamefont {G.~W.}\ \bibnamefont {Wicks}}, \bibinfo {author}
  {\bibfnamefont {S.}~\bibnamefont {Bhattacharjee}},\ and\ \bibinfo {author}
  {\bibfnamefont {A.~N.}\ \bibnamefont {Vamivakas}},\ }\href
  {https://doi.org/10.1038/nphys4303} {\bibfield  {journal} {\bibinfo
  {journal} {Nat. Phys.}\ }\textbf {\bibinfo {volume} {14}},\ \bibinfo {pages}
  {130} (\bibinfo {year} {2017})}\BibitemShut {NoStop}%
\bibitem [{\citenamefont {Emmanuele}\ \emph {et~al.}(2020)\citenamefont
  {Emmanuele}, \citenamefont {Sich}, \citenamefont {Kyriienko}, \citenamefont
  {Shahnazaryan}, \citenamefont {Withers}, \citenamefont {Catanzaro},
  \citenamefont {Walker}, \citenamefont {Benimetskiy}, \citenamefont
  {Skolnick}, \citenamefont {Tartakovskii}, \citenamefont {Shelykh},\ and\
  \citenamefont {Krizhanovskii}}]{Emmanuele:NatCommun11(2020)}%
  \BibitemOpen
  \bibfield  {author} {\bibinfo {author} {\bibfnamefont {R.~P.~A.}\
  \bibnamefont {Emmanuele}}, \bibinfo {author} {\bibfnamefont {M.}~\bibnamefont
  {Sich}}, \bibinfo {author} {\bibfnamefont {O.}~\bibnamefont {Kyriienko}},
  \bibinfo {author} {\bibfnamefont {V.}~\bibnamefont {Shahnazaryan}}, \bibinfo
  {author} {\bibfnamefont {F.}~\bibnamefont {Withers}}, \bibinfo {author}
  {\bibfnamefont {A.}~\bibnamefont {Catanzaro}}, \bibinfo {author}
  {\bibfnamefont {P.~M.}\ \bibnamefont {Walker}}, \bibinfo {author}
  {\bibfnamefont {F.~A.}\ \bibnamefont {Benimetskiy}}, \bibinfo {author}
  {\bibfnamefont {M.~S.}\ \bibnamefont {Skolnick}}, \bibinfo {author}
  {\bibfnamefont {A.~I.}\ \bibnamefont {Tartakovskii}}, \bibinfo {author}
  {\bibfnamefont {I.~A.}\ \bibnamefont {Shelykh}},\ and\ \bibinfo {author}
  {\bibfnamefont {D.~N.}\ \bibnamefont {Krizhanovskii}},\ }\bibfield  {journal}
  {\bibinfo  {journal} {Nat. Commun.}\ }\textbf {\bibinfo {volume} {11}},\
  \href {https://doi.org/10.1038/s41467-020-17340-z}
  {10.1038/s41467-020-17340-z} (\bibinfo {year} {2020})\BibitemShut {NoStop}%
\bibitem [{\citenamefont {Kravtsov}\ \emph {et~al.}(2020)\citenamefont
  {Kravtsov}, \citenamefont {Khestanova}, \citenamefont {Benimetskiy},
  \citenamefont {Ivanova}, \citenamefont {Samusev}, \citenamefont {Sinev},
  \citenamefont {Pidgayko}, \citenamefont {Mozharov}, \citenamefont {Mukhin},
  \citenamefont {Lozhkin}, \citenamefont {Kapitonov}, \citenamefont {Brichkin},
  \citenamefont {Kulakovskii}, \citenamefont {Shelykh}, \citenamefont
  {Tartakovskii}, \citenamefont {Walker}, \citenamefont {Skolnick},
  \citenamefont {Krizhanovskii},\ and\ \citenamefont
  {Iorsh}}]{Kravtsov:LightSci9(2020)}%
  \BibitemOpen
  \bibfield  {author} {\bibinfo {author} {\bibfnamefont {V.}~\bibnamefont
  {Kravtsov}}, \bibinfo {author} {\bibfnamefont {E.}~\bibnamefont
  {Khestanova}}, \bibinfo {author} {\bibfnamefont {F.~A.}\ \bibnamefont
  {Benimetskiy}}, \bibinfo {author} {\bibfnamefont {T.}~\bibnamefont
  {Ivanova}}, \bibinfo {author} {\bibfnamefont {A.~K.}\ \bibnamefont
  {Samusev}}, \bibinfo {author} {\bibfnamefont {I.~S.}\ \bibnamefont {Sinev}},
  \bibinfo {author} {\bibfnamefont {D.}~\bibnamefont {Pidgayko}}, \bibinfo
  {author} {\bibfnamefont {A.~M.}\ \bibnamefont {Mozharov}}, \bibinfo {author}
  {\bibfnamefont {I.~S.}\ \bibnamefont {Mukhin}}, \bibinfo {author}
  {\bibfnamefont {M.~S.}\ \bibnamefont {Lozhkin}}, \bibinfo {author}
  {\bibfnamefont {Y.~V.}\ \bibnamefont {Kapitonov}}, \bibinfo {author}
  {\bibfnamefont {A.~S.}\ \bibnamefont {Brichkin}}, \bibinfo {author}
  {\bibfnamefont {V.~D.}\ \bibnamefont {Kulakovskii}}, \bibinfo {author}
  {\bibfnamefont {I.~A.}\ \bibnamefont {Shelykh}}, \bibinfo {author}
  {\bibfnamefont {A.~I.}\ \bibnamefont {Tartakovskii}}, \bibinfo {author}
  {\bibfnamefont {P.~M.}\ \bibnamefont {Walker}}, \bibinfo {author}
  {\bibfnamefont {M.~S.}\ \bibnamefont {Skolnick}}, \bibinfo {author}
  {\bibfnamefont {D.~N.}\ \bibnamefont {Krizhanovskii}},\ and\ \bibinfo
  {author} {\bibfnamefont {I.~V.}\ \bibnamefont {Iorsh}},\ }\bibfield
  {journal} {\bibinfo  {journal} {Light Sci Appl}\ }\textbf {\bibinfo {volume}
  {9}},\ \href {https://doi.org/10.1038/s41377-020-0286-z}
  {10.1038/s41377-020-0286-z} (\bibinfo {year} {2020})\BibitemShut {NoStop}%
\bibitem [{\citenamefont {Zhang}\ \emph {et~al.}(2018)\citenamefont {Zhang},
  \citenamefont {Gogna}, \citenamefont {Burg}, \citenamefont {Tutuc},\ and\
  \citenamefont {Deng}}]{Zhang2018}%
  \BibitemOpen
  \bibfield  {author} {\bibinfo {author} {\bibfnamefont {L.}~\bibnamefont
  {Zhang}}, \bibinfo {author} {\bibfnamefont {R.}~\bibnamefont {Gogna}},
  \bibinfo {author} {\bibfnamefont {W.}~\bibnamefont {Burg}}, \bibinfo {author}
  {\bibfnamefont {E.}~\bibnamefont {Tutuc}},\ and\ \bibinfo {author}
  {\bibfnamefont {H.}~\bibnamefont {Deng}},\ }\href
  {https://doi.org/10.1038/s41467-018-03188-x} {\bibfield  {journal} {\bibinfo
  {journal} {Nat. Commun.}\ }\textbf {\bibinfo {volume} {9}},\ \bibinfo {pages}
  {713} (\bibinfo {year} {2018})}\BibitemShut {NoStop}%
\bibitem [{\citenamefont {Fernandez}\ \emph {et~al.}(2019)\citenamefont
  {Fernandez}, \citenamefont {Withers}, \citenamefont {Russo},\ and\
  \citenamefont {Barnes}}]{Fernandez2019}%
  \BibitemOpen
  \bibfield  {author} {\bibinfo {author} {\bibfnamefont {H.~A.}\ \bibnamefont
  {Fernandez}}, \bibinfo {author} {\bibfnamefont {F.}~\bibnamefont {Withers}},
  \bibinfo {author} {\bibfnamefont {S.}~\bibnamefont {Russo}},\ and\ \bibinfo
  {author} {\bibfnamefont {W.~L.}\ \bibnamefont {Barnes}},\ }\href
  {https://doi.org/https://doi.org/10.1002/adom.201900484} {\bibfield
  {journal} {\bibinfo  {journal} {Adv. Opt. Mater}\ }\textbf {\bibinfo {volume}
  {7}},\ \bibinfo {pages} {1900484} (\bibinfo {year} {2019})}\BibitemShut
  {NoStop}%
\bibitem [{\citenamefont {Liu}\ \emph {et~al.}(2015)\citenamefont {Liu},
  \citenamefont {Galfsky}, \citenamefont {Sun}, \citenamefont {Xia},
  \citenamefont {Lin}, \citenamefont {Lee}, \citenamefont {K{\'e}na-Cohen},\
  and\ \citenamefont {Menon}}]{Liu2015}%
  \BibitemOpen
  \bibfield  {author} {\bibinfo {author} {\bibfnamefont {X.}~\bibnamefont
  {Liu}}, \bibinfo {author} {\bibfnamefont {T.}~\bibnamefont {Galfsky}},
  \bibinfo {author} {\bibfnamefont {Z.}~\bibnamefont {Sun}}, \bibinfo {author}
  {\bibfnamefont {F.}~\bibnamefont {Xia}}, \bibinfo {author} {\bibfnamefont
  {E.-c.}\ \bibnamefont {Lin}}, \bibinfo {author} {\bibfnamefont {Y.-H.}\
  \bibnamefont {Lee}}, \bibinfo {author} {\bibfnamefont {S.}~\bibnamefont
  {K{\'e}na-Cohen}},\ and\ \bibinfo {author} {\bibfnamefont {V.~M.}\
  \bibnamefont {Menon}},\ }\href {https://doi.org/10.1038/nphoton.2014.304}
  {\bibfield  {journal} {\bibinfo  {journal} {Nature Photonics}\ }\textbf
  {\bibinfo {volume} {9}},\ \bibinfo {pages} {30} (\bibinfo {year}
  {2015})}\BibitemShut {NoStop}%
\bibitem [{\citenamefont {Lackner}\ \emph {et~al.}(2021)\citenamefont
  {Lackner}, \citenamefont {Dusel}, \citenamefont {Egorov}, \citenamefont
  {Han}, \citenamefont {Knopf}, \citenamefont {Eilenberger}, \citenamefont
  {Schr{\"o}der}, \citenamefont {Watanabe}, \citenamefont {Taniguchi},
  \citenamefont {Tongay}, \citenamefont {Anton-Solanas}, \citenamefont
  {H{\"o}fling},\ and\ \citenamefont {Schneider}}]{Lackner2021}%
  \BibitemOpen
  \bibfield  {author} {\bibinfo {author} {\bibfnamefont {L.}~\bibnamefont
  {Lackner}}, \bibinfo {author} {\bibfnamefont {M.}~\bibnamefont {Dusel}},
  \bibinfo {author} {\bibfnamefont {O.~A.}\ \bibnamefont {Egorov}}, \bibinfo
  {author} {\bibfnamefont {B.}~\bibnamefont {Han}}, \bibinfo {author}
  {\bibfnamefont {H.}~\bibnamefont {Knopf}}, \bibinfo {author} {\bibfnamefont
  {F.}~\bibnamefont {Eilenberger}}, \bibinfo {author} {\bibfnamefont
  {S.}~\bibnamefont {Schr{\"o}der}}, \bibinfo {author} {\bibfnamefont
  {K.}~\bibnamefont {Watanabe}}, \bibinfo {author} {\bibfnamefont
  {T.}~\bibnamefont {Taniguchi}}, \bibinfo {author} {\bibfnamefont
  {S.}~\bibnamefont {Tongay}}, \bibinfo {author} {\bibfnamefont
  {C.}~\bibnamefont {Anton-Solanas}}, \bibinfo {author} {\bibfnamefont
  {S.}~\bibnamefont {H{\"o}fling}},\ and\ \bibinfo {author} {\bibfnamefont
  {C.}~\bibnamefont {Schneider}},\ }\href
  {https://doi.org/10.1038/s41467-021-24925-9} {\bibfield  {journal} {\bibinfo
  {journal} {Nat. Comm.}\ }\textbf {\bibinfo {volume} {12}},\ \bibinfo {pages}
  {4933} (\bibinfo {year} {2021})}\BibitemShut {NoStop}%
\bibitem [{\citenamefont {Liu}\ \emph {et~al.}(2016)\citenamefont {Liu},
  \citenamefont {Lee}, \citenamefont {Naylor}, \citenamefont {Ee},
  \citenamefont {Park}, \citenamefont {Johnson},\ and\ \citenamefont
  {Agarwal}}]{Liu:NanoLett16(2016)}%
  \BibitemOpen
  \bibfield  {author} {\bibinfo {author} {\bibfnamefont {W.}~\bibnamefont
  {Liu}}, \bibinfo {author} {\bibfnamefont {B.}~\bibnamefont {Lee}}, \bibinfo
  {author} {\bibfnamefont {C.~H.}\ \bibnamefont {Naylor}}, \bibinfo {author}
  {\bibfnamefont {H.-S.}\ \bibnamefont {Ee}}, \bibinfo {author} {\bibfnamefont
  {J.}~\bibnamefont {Park}}, \bibinfo {author} {\bibfnamefont {A.~T.~C.}\
  \bibnamefont {Johnson}},\ and\ \bibinfo {author} {\bibfnamefont
  {R.}~\bibnamefont {Agarwal}},\ }\href
  {https://doi.org/10.1021/acs.nanolett.5b04588} {\bibfield  {journal}
  {\bibinfo  {journal} {Nano Lett.}\ }\textbf {\bibinfo {volume} {16}},\
  \bibinfo {pages} {1262} (\bibinfo {year} {2016})}\BibitemShut {NoStop}%
\bibitem [{\citenamefont {Wen}\ \emph {et~al.}(2017)\citenamefont {Wen},
  \citenamefont {Wang}, \citenamefont {Wang}, \citenamefont {Deng},
  \citenamefont {Zhuang}, \citenamefont {Zhang}, \citenamefont {Liu},
  \citenamefont {She}, \citenamefont {Chen}, \citenamefont {Chen},
  \citenamefont {Deng},\ and\ \citenamefont {Xu}}]{Wen:NanoLett17(2017)}%
  \BibitemOpen
  \bibfield  {author} {\bibinfo {author} {\bibfnamefont {J.}~\bibnamefont
  {Wen}}, \bibinfo {author} {\bibfnamefont {H.}~\bibnamefont {Wang}}, \bibinfo
  {author} {\bibfnamefont {W.}~\bibnamefont {Wang}}, \bibinfo {author}
  {\bibfnamefont {Z.}~\bibnamefont {Deng}}, \bibinfo {author} {\bibfnamefont
  {C.}~\bibnamefont {Zhuang}}, \bibinfo {author} {\bibfnamefont
  {Y.}~\bibnamefont {Zhang}}, \bibinfo {author} {\bibfnamefont
  {F.}~\bibnamefont {Liu}}, \bibinfo {author} {\bibfnamefont {J.}~\bibnamefont
  {She}}, \bibinfo {author} {\bibfnamefont {J.}~\bibnamefont {Chen}}, \bibinfo
  {author} {\bibfnamefont {H.}~\bibnamefont {Chen}}, \bibinfo {author}
  {\bibfnamefont {S.}~\bibnamefont {Deng}},\ and\ \bibinfo {author}
  {\bibfnamefont {N.}~\bibnamefont {Xu}},\ }\href
  {https://doi.org/10.1021/acs.nanolett.7b01344} {\bibfield  {journal}
  {\bibinfo  {journal} {Nano Lett.}\ }\textbf {\bibinfo {volume} {17}},\
  \bibinfo {pages} {4689} (\bibinfo {year} {2017})}\BibitemShut {NoStop}%
\bibitem [{\citenamefont {Abid}\ \emph {et~al.}(2017)\citenamefont {Abid},
  \citenamefont {Chen}, \citenamefont {Yuan}, \citenamefont {Bohloul},
  \citenamefont {Najmaei}, \citenamefont {Avendano}, \citenamefont
  {P{\'{e}}chou}, \citenamefont {Mlayah},\ and\ \citenamefont
  {Lou}}]{Abid:ACSPhoto4(2017)}%
  \BibitemOpen
  \bibfield  {author} {\bibinfo {author} {\bibfnamefont {I.}~\bibnamefont
  {Abid}}, \bibinfo {author} {\bibfnamefont {W.}~\bibnamefont {Chen}}, \bibinfo
  {author} {\bibfnamefont {J.}~\bibnamefont {Yuan}}, \bibinfo {author}
  {\bibfnamefont {A.}~\bibnamefont {Bohloul}}, \bibinfo {author} {\bibfnamefont
  {S.}~\bibnamefont {Najmaei}}, \bibinfo {author} {\bibfnamefont
  {C.}~\bibnamefont {Avendano}}, \bibinfo {author} {\bibfnamefont
  {R.}~\bibnamefont {P{\'{e}}chou}}, \bibinfo {author} {\bibfnamefont
  {A.}~\bibnamefont {Mlayah}},\ and\ \bibinfo {author} {\bibfnamefont
  {J.}~\bibnamefont {Lou}},\ }\href
  {https://doi.org/10.1021/acsphotonics.6b00957} {\bibfield  {journal}
  {\bibinfo  {journal} {{ACS} Photonics}\ }\textbf {\bibinfo {volume} {4}},\
  \bibinfo {pages} {1653} (\bibinfo {year} {2017})}\BibitemShut {NoStop}%
\bibitem [{\citenamefont {St{\"u}hrenberg}\ \emph {et~al.}(2018)\citenamefont
  {St{\"u}hrenberg}, \citenamefont {Munkhbat}, \citenamefont {Baranov},
  \citenamefont {Cuadra}, \citenamefont {Yankovich}, \citenamefont
  {Antosiewicz}, \citenamefont {Olsson},\ and\ \citenamefont
  {Shegai}}]{Stuhrenberg2018}%
  \BibitemOpen
  \bibfield  {author} {\bibinfo {author} {\bibfnamefont {M.}~\bibnamefont
  {St{\"u}hrenberg}}, \bibinfo {author} {\bibfnamefont {B.}~\bibnamefont
  {Munkhbat}}, \bibinfo {author} {\bibfnamefont {D.~G.}\ \bibnamefont
  {Baranov}}, \bibinfo {author} {\bibfnamefont {J.}~\bibnamefont {Cuadra}},
  \bibinfo {author} {\bibfnamefont {A.~B.}\ \bibnamefont {Yankovich}}, \bibinfo
  {author} {\bibfnamefont {T.~J.}\ \bibnamefont {Antosiewicz}}, \bibinfo
  {author} {\bibfnamefont {E.}~\bibnamefont {Olsson}},\ and\ \bibinfo {author}
  {\bibfnamefont {T.}~\bibnamefont {Shegai}},\ }\href
  {https://doi.org/10.1021/acs.nanolett.8b02652} {\bibfield  {journal}
  {\bibinfo  {journal} {Nano Lett.}\ }\textbf {\bibinfo {volume} {18}},\
  \bibinfo {pages} {5938} (\bibinfo {year} {2018})}\BibitemShut {NoStop}%
\bibitem [{\citenamefont {Yankovich}\ \emph {et~al.}(2019)\citenamefont
  {Yankovich}, \citenamefont {Munkhbat}, \citenamefont {Baranov}, \citenamefont
  {Cuadra}, \citenamefont {Ols{\'e}n}, \citenamefont {Louren{\c{c}}o-Martins},
  \citenamefont {Tizei}, \citenamefont {Kociak}, \citenamefont {Olsson},\ and\
  \citenamefont {Shegai}}]{Yankovich2019}%
  \BibitemOpen
  \bibfield  {author} {\bibinfo {author} {\bibfnamefont {A.~B.}\ \bibnamefont
  {Yankovich}}, \bibinfo {author} {\bibfnamefont {B.}~\bibnamefont {Munkhbat}},
  \bibinfo {author} {\bibfnamefont {D.~G.}\ \bibnamefont {Baranov}}, \bibinfo
  {author} {\bibfnamefont {J.}~\bibnamefont {Cuadra}}, \bibinfo {author}
  {\bibfnamefont {E.}~\bibnamefont {Ols{\'e}n}}, \bibinfo {author}
  {\bibfnamefont {H.}~\bibnamefont {Louren{\c{c}}o-Martins}}, \bibinfo {author}
  {\bibfnamefont {L.~H.~G.}\ \bibnamefont {Tizei}}, \bibinfo {author}
  {\bibfnamefont {M.}~\bibnamefont {Kociak}}, \bibinfo {author} {\bibfnamefont
  {E.}~\bibnamefont {Olsson}},\ and\ \bibinfo {author} {\bibfnamefont
  {T.}~\bibnamefont {Shegai}},\ }\href
  {https://doi.org/10.1021/acs.nanolett.9b03534} {\bibfield  {journal}
  {\bibinfo  {journal} {Nano Lett.}\ }\textbf {\bibinfo {volume} {19}},\
  \bibinfo {pages} {8171} (\bibinfo {year} {2019})}\BibitemShut {NoStop}%
\bibitem [{\citenamefont {Gonçalves}\ \emph {et~al.}(2020)\citenamefont
  {Gonçalves}, \citenamefont {Stenger}, \citenamefont {Cox}, \citenamefont
  {Mortensen},\ and\ \citenamefont {Xiao}}]{Goncalves2020}%
  \BibitemOpen
  \bibfield  {author} {\bibinfo {author} {\bibfnamefont {P.~A.~D.}\
  \bibnamefont {Gonçalves}}, \bibinfo {author} {\bibfnamefont
  {N.}~\bibnamefont {Stenger}}, \bibinfo {author} {\bibfnamefont {J.~D.}\
  \bibnamefont {Cox}}, \bibinfo {author} {\bibfnamefont {N.~A.}\ \bibnamefont
  {Mortensen}},\ and\ \bibinfo {author} {\bibfnamefont {S.}~\bibnamefont
  {Xiao}},\ }\href {https://doi.org/https://doi.org/10.1002/adom.201901473}
  {\bibfield  {journal} {\bibinfo  {journal} {Adv. Opt. Mater.}\ }\textbf
  {\bibinfo {volume} {8}},\ \bibinfo {pages} {1901473} (\bibinfo {year}
  {2020})}\BibitemShut {NoStop}%
\bibitem [{\citenamefont {Schneider}\ \emph {et~al.}(2018)\citenamefont
  {Schneider}, \citenamefont {Glazov}, \citenamefont {Korn}, \citenamefont
  {H{\"o}fling},\ and\ \citenamefont {Urbaszek}}]{Schneider2018}%
  \BibitemOpen
  \bibfield  {author} {\bibinfo {author} {\bibfnamefont {C.}~\bibnamefont
  {Schneider}}, \bibinfo {author} {\bibfnamefont {M.~M.}\ \bibnamefont
  {Glazov}}, \bibinfo {author} {\bibfnamefont {T.}~\bibnamefont {Korn}},
  \bibinfo {author} {\bibfnamefont {S.}~\bibnamefont {H{\"o}fling}},\ and\
  \bibinfo {author} {\bibfnamefont {B.}~\bibnamefont {Urbaszek}},\ }\href
  {https://doi.org/10.1038/s41467-018-04866-6} {\bibfield  {journal} {\bibinfo
  {journal} {Nat. Comm.}\ }\textbf {\bibinfo {volume} {9}},\ \bibinfo {pages}
  {2695} (\bibinfo {year} {2018})}\BibitemShut {NoStop}%
\bibitem [{\citenamefont {Huang}\ \emph {et~al.}(2022)\citenamefont {Huang},
  \citenamefont {Krasnok}, \citenamefont {Al{\'{u}}}, \citenamefont {Yu},
  \citenamefont {Neshev},\ and\ \citenamefont {Miroshnichenko}}]{Huang_2022}%
  \BibitemOpen
  \bibfield  {author} {\bibinfo {author} {\bibfnamefont {L.}~\bibnamefont
  {Huang}}, \bibinfo {author} {\bibfnamefont {A.}~\bibnamefont {Krasnok}},
  \bibinfo {author} {\bibfnamefont {A.}~\bibnamefont {Al{\'{u}}}}, \bibinfo
  {author} {\bibfnamefont {Y.}~\bibnamefont {Yu}}, \bibinfo {author}
  {\bibfnamefont {D.}~\bibnamefont {Neshev}},\ and\ \bibinfo {author}
  {\bibfnamefont {A.~E.}\ \bibnamefont {Miroshnichenko}},\ }\href
  {https://doi.org/10.1088/1361-6633/ac45f9} {\bibfield  {journal} {\bibinfo
  {journal} {Reports on Progress in Physics}\ }\textbf {\bibinfo {volume}
  {85}},\ \bibinfo {pages} {046401} (\bibinfo {year} {2022})}\BibitemShut
  {NoStop}%
\bibitem [{\citenamefont {Carusotto}\ and\ \citenamefont
  {Ciuti}(2013)}]{CarusottoCiuti2013}%
  \BibitemOpen
  \bibfield  {author} {\bibinfo {author} {\bibfnamefont {I.}~\bibnamefont
  {Carusotto}}\ and\ \bibinfo {author} {\bibfnamefont {C.}~\bibnamefont
  {Ciuti}},\ }\href {https://doi.org/10.1103/RevModPhys.85.299} {\bibfield
  {journal} {\bibinfo  {journal} {Rev. Mod. Phys.}\ }\textbf {\bibinfo {volume}
  {85}},\ \bibinfo {pages} {299} (\bibinfo {year} {2013})}\BibitemShut
  {NoStop}%
\bibitem [{\citenamefont {Basov}\ \emph {et~al.}(2021)\citenamefont {Basov},
  \citenamefont {Asenjo-Garcia}, \citenamefont {Schuck}, \citenamefont {Zhu},\
  and\ \citenamefont {Rubio}}]{BasovAsenjo2021}%
  \BibitemOpen
  \bibfield  {author} {\bibinfo {author} {\bibfnamefont {D.~N.}\ \bibnamefont
  {Basov}}, \bibinfo {author} {\bibfnamefont {A.}~\bibnamefont
  {Asenjo-Garcia}}, \bibinfo {author} {\bibfnamefont {P.~J.}\ \bibnamefont
  {Schuck}}, \bibinfo {author} {\bibfnamefont {X.}~\bibnamefont {Zhu}},\ and\
  \bibinfo {author} {\bibfnamefont {A.}~\bibnamefont {Rubio}},\ }\href
  {https://doi.org/doi:10.1515/nanoph-2020-0449} {\bibfield  {journal}
  {\bibinfo  {journal} {Nanophotonics}\ }\textbf {\bibinfo {volume} {10}},\
  \bibinfo {pages} {549} (\bibinfo {year} {2021})}\BibitemShut {NoStop}%
\bibitem [{\citenamefont {Shahnazaryan}\ \emph {et~al.}(2017)\citenamefont
  {Shahnazaryan}, \citenamefont {Iorsh}, \citenamefont {Shelykh},\ and\
  \citenamefont {Kyriienko}}]{Shahnazaryan:PRB96(2017)}%
  \BibitemOpen
  \bibfield  {author} {\bibinfo {author} {\bibfnamefont {V.}~\bibnamefont
  {Shahnazaryan}}, \bibinfo {author} {\bibfnamefont {I.}~\bibnamefont {Iorsh}},
  \bibinfo {author} {\bibfnamefont {I.~A.}\ \bibnamefont {Shelykh}},\ and\
  \bibinfo {author} {\bibfnamefont {O.}~\bibnamefont {Kyriienko}},\ }\href
  {https://doi.org/10.1103/physrevb.96.115409} {\bibfield  {journal} {\bibinfo
  {journal} {Phys. Rev. B}\ }\textbf {\bibinfo {volume} {96}},\ \bibinfo
  {pages} {115409} (\bibinfo {year} {2017})}\BibitemShut {NoStop}%
\bibitem [{\citenamefont {Barachati}\ \emph {et~al.}(2018)\citenamefont
  {Barachati}, \citenamefont {Fieramosca}, \citenamefont {Hafezian},
  \citenamefont {Gu}, \citenamefont {Chakraborty}, \citenamefont {Ballarini},
  \citenamefont {Martinu}, \citenamefont {Menon}, \citenamefont {Sanvitto},\
  and\ \citenamefont {K{\'{e}}na-Cohen}}]{Barachati:NatNano13(2018)}%
  \BibitemOpen
  \bibfield  {author} {\bibinfo {author} {\bibfnamefont {F.}~\bibnamefont
  {Barachati}}, \bibinfo {author} {\bibfnamefont {A.}~\bibnamefont
  {Fieramosca}}, \bibinfo {author} {\bibfnamefont {S.}~\bibnamefont
  {Hafezian}}, \bibinfo {author} {\bibfnamefont {J.}~\bibnamefont {Gu}},
  \bibinfo {author} {\bibfnamefont {B.}~\bibnamefont {Chakraborty}}, \bibinfo
  {author} {\bibfnamefont {D.}~\bibnamefont {Ballarini}}, \bibinfo {author}
  {\bibfnamefont {L.}~\bibnamefont {Martinu}}, \bibinfo {author} {\bibfnamefont
  {V.}~\bibnamefont {Menon}}, \bibinfo {author} {\bibfnamefont
  {D.}~\bibnamefont {Sanvitto}},\ and\ \bibinfo {author} {\bibfnamefont
  {S.}~\bibnamefont {K{\'{e}}na-Cohen}},\ }\href
  {https://doi.org/10.1038/s41565-018-0219-7} {\bibfield  {journal} {\bibinfo
  {journal} {Nat. Nanotechnol.}\ }\textbf {\bibinfo {volume} {13}},\ \bibinfo
  {pages} {906} (\bibinfo {year} {2018})}\BibitemShut {NoStop}%
\bibitem [{\citenamefont {Bleu}\ \emph {et~al.}(2020)\citenamefont {Bleu},
  \citenamefont {Li}, \citenamefont {Levinsen},\ and\ \citenamefont
  {Parish}}]{Bleu2020}%
  \BibitemOpen
  \bibfield  {author} {\bibinfo {author} {\bibfnamefont {O.}~\bibnamefont
  {Bleu}}, \bibinfo {author} {\bibfnamefont {G.}~\bibnamefont {Li}}, \bibinfo
  {author} {\bibfnamefont {J.}~\bibnamefont {Levinsen}},\ and\ \bibinfo
  {author} {\bibfnamefont {M.~M.}\ \bibnamefont {Parish}},\ }\href
  {https://doi.org/10.1103/PhysRevResearch.2.043185} {\bibfield  {journal}
  {\bibinfo  {journal} {Phys. Rev. Research}\ }\textbf {\bibinfo {volume}
  {2}},\ \bibinfo {pages} {043185} (\bibinfo {year} {2020})}\BibitemShut
  {NoStop}%
\bibitem [{\citenamefont {Sim}\ \emph {et~al.}(2020)\citenamefont {Sim},
  \citenamefont {Lee}, \citenamefont {Lee}, \citenamefont {Cha}, \citenamefont
  {Cha}, \citenamefont {Heo}, \citenamefont {Cho}, \citenamefont {Shim},
  \citenamefont {Lee}, \citenamefont {Yoo}, \citenamefont {Prasankumar},
  \citenamefont {Choi},\ and\ \citenamefont {Jo}}]{SimPRB2020}%
  \BibitemOpen
  \bibfield  {author} {\bibinfo {author} {\bibfnamefont {S.}~\bibnamefont
  {Sim}}, \bibinfo {author} {\bibfnamefont {D.}~\bibnamefont {Lee}}, \bibinfo
  {author} {\bibfnamefont {J.}~\bibnamefont {Lee}}, \bibinfo {author}
  {\bibfnamefont {M.}~\bibnamefont {Cha}}, \bibinfo {author} {\bibfnamefont
  {S.}~\bibnamefont {Cha}}, \bibinfo {author} {\bibfnamefont {W.}~\bibnamefont
  {Heo}}, \bibinfo {author} {\bibfnamefont {S.}~\bibnamefont {Cho}}, \bibinfo
  {author} {\bibfnamefont {W.}~\bibnamefont {Shim}}, \bibinfo {author}
  {\bibfnamefont {K.}~\bibnamefont {Lee}}, \bibinfo {author} {\bibfnamefont
  {J.}~\bibnamefont {Yoo}}, \bibinfo {author} {\bibfnamefont {R.~P.}\
  \bibnamefont {Prasankumar}}, \bibinfo {author} {\bibfnamefont
  {H.}~\bibnamefont {Choi}},\ and\ \bibinfo {author} {\bibfnamefont {M.-H.}\
  \bibnamefont {Jo}},\ }\href {https://doi.org/10.1103/PhysRevB.101.174309}
  {\bibfield  {journal} {\bibinfo  {journal} {Phys. Rev. B}\ }\textbf {\bibinfo
  {volume} {101}},\ \bibinfo {pages} {174309} (\bibinfo {year}
  {2020})}\BibitemShut {NoStop}%
\bibitem [{\citenamefont {Gu}\ \emph {et~al.}(2021)\citenamefont {Gu},
  \citenamefont {Walther}, \citenamefont {Waldecker}, \citenamefont {Rhodes},
  \citenamefont {Raja}, \citenamefont {Hone}, \citenamefont {Heinz},
  \citenamefont {K{\'e}na-Cohen}, \citenamefont {Pohl},\ and\ \citenamefont
  {Menon}}]{Gu2021}%
  \BibitemOpen
  \bibfield  {author} {\bibinfo {author} {\bibfnamefont {J.}~\bibnamefont
  {Gu}}, \bibinfo {author} {\bibfnamefont {V.}~\bibnamefont {Walther}},
  \bibinfo {author} {\bibfnamefont {L.}~\bibnamefont {Waldecker}}, \bibinfo
  {author} {\bibfnamefont {D.}~\bibnamefont {Rhodes}}, \bibinfo {author}
  {\bibfnamefont {A.}~\bibnamefont {Raja}}, \bibinfo {author} {\bibfnamefont
  {J.~C.}\ \bibnamefont {Hone}}, \bibinfo {author} {\bibfnamefont {T.~F.}\
  \bibnamefont {Heinz}}, \bibinfo {author} {\bibfnamefont {S.}~\bibnamefont
  {K{\'e}na-Cohen}}, \bibinfo {author} {\bibfnamefont {T.}~\bibnamefont
  {Pohl}},\ and\ \bibinfo {author} {\bibfnamefont {V.~M.}\ \bibnamefont
  {Menon}},\ }\href {https://doi.org/10.1038/s41467-021-22537-x} {\bibfield
  {journal} {\bibinfo  {journal} {Nat. Commun.}\ }\textbf {\bibinfo {volume}
  {12}},\ \bibinfo {pages} {2269} (\bibinfo {year} {2021})}\BibitemShut
  {NoStop}%
\bibitem [{\citenamefont {Erkensten}\ \emph {et~al.}(2021)\citenamefont
  {Erkensten}, \citenamefont {Brem},\ and\ \citenamefont
  {Malic}}]{Erkensten2021}%
  \BibitemOpen
  \bibfield  {author} {\bibinfo {author} {\bibfnamefont {D.}~\bibnamefont
  {Erkensten}}, \bibinfo {author} {\bibfnamefont {S.}~\bibnamefont {Brem}},\
  and\ \bibinfo {author} {\bibfnamefont {E.}~\bibnamefont {Malic}},\ }\href
  {https://doi.org/10.1103/PhysRevB.103.045426} {\bibfield  {journal} {\bibinfo
   {journal} {Phys. Rev. B}\ }\textbf {\bibinfo {volume} {103}},\ \bibinfo
  {pages} {045426} (\bibinfo {year} {2021})}\BibitemShut {NoStop}%
\bibitem [{\citenamefont {Stepanov}\ \emph {et~al.}(2021)\citenamefont
  {Stepanov}, \citenamefont {Vashisht}, \citenamefont {Klaas}, \citenamefont
  {Lundt}, \citenamefont {Tongay}, \citenamefont {Blei}, \citenamefont
  {Höfling}, \citenamefont {Volz}, \citenamefont {Minguzzi}, \citenamefont
  {Renard}, \citenamefont {Schneider},\ and\ \citenamefont
  {Richard}}]{Stepanov:PRL126(2021)}%
  \BibitemOpen
  \bibfield  {author} {\bibinfo {author} {\bibfnamefont {P.}~\bibnamefont
  {Stepanov}}, \bibinfo {author} {\bibfnamefont {A.}~\bibnamefont {Vashisht}},
  \bibinfo {author} {\bibfnamefont {M.}~\bibnamefont {Klaas}}, \bibinfo
  {author} {\bibfnamefont {N.}~\bibnamefont {Lundt}}, \bibinfo {author}
  {\bibfnamefont {S.}~\bibnamefont {Tongay}}, \bibinfo {author} {\bibfnamefont
  {M.}~\bibnamefont {Blei}}, \bibinfo {author} {\bibfnamefont {S.}~\bibnamefont
  {Höfling}}, \bibinfo {author} {\bibfnamefont {T.}~\bibnamefont {Volz}},
  \bibinfo {author} {\bibfnamefont {A.}~\bibnamefont {Minguzzi}}, \bibinfo
  {author} {\bibfnamefont {J.}~\bibnamefont {Renard}}, \bibinfo {author}
  {\bibfnamefont {C.}~\bibnamefont {Schneider}},\ and\ \bibinfo {author}
  {\bibfnamefont {M.}~\bibnamefont {Richard}},\ }\href
  {https://doi.org/10.1103/physrevlett.126.167401} {\bibfield  {journal}
  {\bibinfo  {journal} {Phys. Rev. Lett.}\ }\textbf {\bibinfo {volume} {126}},\
  \bibinfo {pages} {167401} (\bibinfo {year} {2021})}\BibitemShut {NoStop}%
\bibitem [{\citenamefont {Anton-Solanas}\ \emph {et~al.}(2021)\citenamefont
  {Anton-Solanas}, \citenamefont {Waldherr}, \citenamefont {Klaas},
  \citenamefont {Suchomel}, \citenamefont {Harder}, \citenamefont {Cai},
  \citenamefont {Sedov}, \citenamefont {Klembt}, \citenamefont {Kavokin},
  \citenamefont {Tongay}, \citenamefont {Watanabe}, \citenamefont {Taniguchi},
  \citenamefont {H{\"o}fling},\ and\ \citenamefont
  {Schneider}}]{Anton-Solanas2021}%
  \BibitemOpen
  \bibfield  {author} {\bibinfo {author} {\bibfnamefont {C.}~\bibnamefont
  {Anton-Solanas}}, \bibinfo {author} {\bibfnamefont {M.}~\bibnamefont
  {Waldherr}}, \bibinfo {author} {\bibfnamefont {M.}~\bibnamefont {Klaas}},
  \bibinfo {author} {\bibfnamefont {H.}~\bibnamefont {Suchomel}}, \bibinfo
  {author} {\bibfnamefont {T.~H.}\ \bibnamefont {Harder}}, \bibinfo {author}
  {\bibfnamefont {H.}~\bibnamefont {Cai}}, \bibinfo {author} {\bibfnamefont
  {E.}~\bibnamefont {Sedov}}, \bibinfo {author} {\bibfnamefont
  {S.}~\bibnamefont {Klembt}}, \bibinfo {author} {\bibfnamefont {A.~V.}\
  \bibnamefont {Kavokin}}, \bibinfo {author} {\bibfnamefont {S.}~\bibnamefont
  {Tongay}}, \bibinfo {author} {\bibfnamefont {K.}~\bibnamefont {Watanabe}},
  \bibinfo {author} {\bibfnamefont {T.}~\bibnamefont {Taniguchi}}, \bibinfo
  {author} {\bibfnamefont {S.}~\bibnamefont {H{\"o}fling}},\ and\ \bibinfo
  {author} {\bibfnamefont {C.}~\bibnamefont {Schneider}},\ }\href
  {https://doi.org/10.1038/s41563-021-01000-8} {\bibfield  {journal} {\bibinfo
  {journal} {Nat. Mater.}\ }\textbf {\bibinfo {volume} {20}},\ \bibinfo {pages}
  {1233} (\bibinfo {year} {2021})}\BibitemShut {NoStop}%
\bibitem [{\citenamefont {Mak}\ \emph {et~al.}(2012)\citenamefont {Mak},
  \citenamefont {He}, \citenamefont {Lee}, \citenamefont {Lee}, \citenamefont
  {Hone}, \citenamefont {Heinz},\ and\ \citenamefont
  {Shan}}]{Mak:NatMat12(2012)}%
  \BibitemOpen
  \bibfield  {author} {\bibinfo {author} {\bibfnamefont {K.~F.}\ \bibnamefont
  {Mak}}, \bibinfo {author} {\bibfnamefont {K.}~\bibnamefont {He}}, \bibinfo
  {author} {\bibfnamefont {C.}~\bibnamefont {Lee}}, \bibinfo {author}
  {\bibfnamefont {G.~H.}\ \bibnamefont {Lee}}, \bibinfo {author} {\bibfnamefont
  {J.}~\bibnamefont {Hone}}, \bibinfo {author} {\bibfnamefont {T.~F.}\
  \bibnamefont {Heinz}},\ and\ \bibinfo {author} {\bibfnamefont
  {J.}~\bibnamefont {Shan}},\ }\href {https://doi.org/10.1038/nmat3505}
  {\bibfield  {journal} {\bibinfo  {journal} {Nat. Mater.}\ }\textbf {\bibinfo
  {volume} {12}},\ \bibinfo {pages} {207} (\bibinfo {year} {2012})}\BibitemShut
  {NoStop}%
\bibitem [{\citenamefont {Ross}\ \emph {et~al.}(2013)\citenamefont {Ross},
  \citenamefont {Wu}, \citenamefont {Yu}, \citenamefont {Ghimire},
  \citenamefont {Jones}, \citenamefont {Aivazian}, \citenamefont {Yan},
  \citenamefont {Mandrus}, \citenamefont {Xiao}, \citenamefont {Yao},\ and\
  \citenamefont {Xu}}]{Ross:NatComm4(2013)}%
  \BibitemOpen
  \bibfield  {author} {\bibinfo {author} {\bibfnamefont {J.~S.}\ \bibnamefont
  {Ross}}, \bibinfo {author} {\bibfnamefont {S.}~\bibnamefont {Wu}}, \bibinfo
  {author} {\bibfnamefont {H.}~\bibnamefont {Yu}}, \bibinfo {author}
  {\bibfnamefont {N.~J.}\ \bibnamefont {Ghimire}}, \bibinfo {author}
  {\bibfnamefont {A.~M.}\ \bibnamefont {Jones}}, \bibinfo {author}
  {\bibfnamefont {G.}~\bibnamefont {Aivazian}}, \bibinfo {author}
  {\bibfnamefont {J.}~\bibnamefont {Yan}}, \bibinfo {author} {\bibfnamefont
  {D.~G.}\ \bibnamefont {Mandrus}}, \bibinfo {author} {\bibfnamefont
  {D.}~\bibnamefont {Xiao}}, \bibinfo {author} {\bibfnamefont {W.}~\bibnamefont
  {Yao}},\ and\ \bibinfo {author} {\bibfnamefont {X.}~\bibnamefont {Xu}},\
  }\bibfield  {journal} {\bibinfo  {journal} {Nat. Commun.}\ }\textbf {\bibinfo
  {volume} {4}},\ \href {https://doi.org/10.1038/ncomms2498}
  {10.1038/ncomms2498} (\bibinfo {year} {2013})\BibitemShut {NoStop}%
\bibitem [{\citenamefont {Lyons}\ \emph {et~al.}(2019)\citenamefont {Lyons},
  \citenamefont {Dufferwiel}, \citenamefont {Brooks}, \citenamefont {Withers},
  \citenamefont {Taniguchi}, \citenamefont {Watanabe}, \citenamefont
  {Novoselov}, \citenamefont {Burkard},\ and\ \citenamefont
  {Tartakovskii}}]{Lyons2019}%
  \BibitemOpen
  \bibfield  {author} {\bibinfo {author} {\bibfnamefont {T.~P.}\ \bibnamefont
  {Lyons}}, \bibinfo {author} {\bibfnamefont {S.}~\bibnamefont {Dufferwiel}},
  \bibinfo {author} {\bibfnamefont {M.}~\bibnamefont {Brooks}}, \bibinfo
  {author} {\bibfnamefont {F.}~\bibnamefont {Withers}}, \bibinfo {author}
  {\bibfnamefont {T.}~\bibnamefont {Taniguchi}}, \bibinfo {author}
  {\bibfnamefont {K.}~\bibnamefont {Watanabe}}, \bibinfo {author}
  {\bibfnamefont {K.~S.}\ \bibnamefont {Novoselov}}, \bibinfo {author}
  {\bibfnamefont {G.}~\bibnamefont {Burkard}},\ and\ \bibinfo {author}
  {\bibfnamefont {A.~I.}\ \bibnamefont {Tartakovskii}},\ }\href
  {https://doi.org/10.1038/s41467-019-10228-7} {\bibfield  {journal} {\bibinfo
  {journal} {Nat. Comm.}\ }\textbf {\bibinfo {volume} {10}},\ \bibinfo {pages}
  {2330} (\bibinfo {year} {2019})},\ \bibinfo {note}
  {31133703[pmid]}\BibitemShut {NoStop}%
\bibitem [{\citenamefont {Zipfel}\ \emph {et~al.}(2020)\citenamefont {Zipfel},
  \citenamefont {Wagner}, \citenamefont {Ziegler}, \citenamefont {Taniguchi},
  \citenamefont {Watanabe}, \citenamefont {Semina},\ and\ \citenamefont
  {Chernikov}}]{Zipfel2020}%
  \BibitemOpen
  \bibfield  {author} {\bibinfo {author} {\bibfnamefont {J.}~\bibnamefont
  {Zipfel}}, \bibinfo {author} {\bibfnamefont {K.}~\bibnamefont {Wagner}},
  \bibinfo {author} {\bibfnamefont {J.~D.}\ \bibnamefont {Ziegler}}, \bibinfo
  {author} {\bibfnamefont {T.}~\bibnamefont {Taniguchi}}, \bibinfo {author}
  {\bibfnamefont {K.}~\bibnamefont {Watanabe}}, \bibinfo {author}
  {\bibfnamefont {M.~A.}\ \bibnamefont {Semina}},\ and\ \bibinfo {author}
  {\bibfnamefont {A.}~\bibnamefont {Chernikov}},\ }\href
  {https://doi.org/10.1063/5.0012721} {\bibfield  {journal} {\bibinfo
  {journal} {The Journal of Chemical Physics}\ }\textbf {\bibinfo {volume}
  {153}},\ \bibinfo {pages} {034706} (\bibinfo {year} {2020})},\ \Eprint
  {https://arxiv.org/abs/https://doi.org/10.1063/5.0012721}
  {https://doi.org/10.1063/5.0012721} \BibitemShut {NoStop}%
\bibitem [{\citenamefont {Zipfel}\ \emph {et~al.}(2022)\citenamefont {Zipfel},
  \citenamefont {Wagner}, \citenamefont {Semina}, \citenamefont {Ziegler},
  \citenamefont {Taniguchi}, \citenamefont {Watanabe}, \citenamefont {Glazov},\
  and\ \citenamefont {Chernikov}}]{Zipfel2022}%
  \BibitemOpen
  \bibfield  {author} {\bibinfo {author} {\bibfnamefont {J.}~\bibnamefont
  {Zipfel}}, \bibinfo {author} {\bibfnamefont {K.}~\bibnamefont {Wagner}},
  \bibinfo {author} {\bibfnamefont {M.~A.}\ \bibnamefont {Semina}}, \bibinfo
  {author} {\bibfnamefont {J.~D.}\ \bibnamefont {Ziegler}}, \bibinfo {author}
  {\bibfnamefont {T.}~\bibnamefont {Taniguchi}}, \bibinfo {author}
  {\bibfnamefont {K.}~\bibnamefont {Watanabe}}, \bibinfo {author}
  {\bibfnamefont {M.~M.}\ \bibnamefont {Glazov}},\ and\ \bibinfo {author}
  {\bibfnamefont {A.}~\bibnamefont {Chernikov}},\ }\href
  {https://doi.org/10.1103/PhysRevB.105.075311} {\bibfield  {journal} {\bibinfo
   {journal} {Phys. Rev. B}\ }\textbf {\bibinfo {volume} {105}},\ \bibinfo
  {pages} {075311} (\bibinfo {year} {2022})}\BibitemShut {NoStop}%
\bibitem [{\citenamefont {Finkelstein}\ \emph {et~al.}(1995)\citenamefont
  {Finkelstein}, \citenamefont {Shtrikman},\ and\ \citenamefont
  {Bar-Joseph}}]{Finkelstein:PRL74(1995)}%
  \BibitemOpen
  \bibfield  {author} {\bibinfo {author} {\bibfnamefont {G.}~\bibnamefont
  {Finkelstein}}, \bibinfo {author} {\bibfnamefont {H.}~\bibnamefont
  {Shtrikman}},\ and\ \bibinfo {author} {\bibfnamefont {I.}~\bibnamefont
  {Bar-Joseph}},\ }\href {https://doi.org/10.1103/physrevlett.74.976}
  {\bibfield  {journal} {\bibinfo  {journal} {Phys. Rev. Lett.}\ }\textbf
  {\bibinfo {volume} {74}},\ \bibinfo {pages} {976} (\bibinfo {year}
  {1995})}\BibitemShut {NoStop}%
\bibitem [{\citenamefont {Glasberg}\ \emph {et~al.}(1999)\citenamefont
  {Glasberg}, \citenamefont {Finkelstein}, \citenamefont {Shtrikman},\ and\
  \citenamefont {Bar-Joseph}}]{Glasberg1999}%
  \BibitemOpen
  \bibfield  {author} {\bibinfo {author} {\bibfnamefont {S.}~\bibnamefont
  {Glasberg}}, \bibinfo {author} {\bibfnamefont {G.}~\bibnamefont
  {Finkelstein}}, \bibinfo {author} {\bibfnamefont {H.}~\bibnamefont
  {Shtrikman}},\ and\ \bibinfo {author} {\bibfnamefont {I.}~\bibnamefont
  {Bar-Joseph}},\ }\href {https://doi.org/10.1103/PhysRevB.59.R10425}
  {\bibfield  {journal} {\bibinfo  {journal} {Phys. Rev. B}\ }\textbf {\bibinfo
  {volume} {59}},\ \bibinfo {pages} {R10425} (\bibinfo {year}
  {1999})}\BibitemShut {NoStop}%
\bibitem [{\citenamefont {Rapaport}\ \emph {et~al.}(2000)\citenamefont
  {Rapaport}, \citenamefont {Harel}, \citenamefont {Cohen}, \citenamefont
  {Ron}, \citenamefont {Linder},\ and\ \citenamefont
  {Pfeiffer}}]{Rapaport:PRL84(2000)}%
  \BibitemOpen
  \bibfield  {author} {\bibinfo {author} {\bibfnamefont {R.}~\bibnamefont
  {Rapaport}}, \bibinfo {author} {\bibfnamefont {R.}~\bibnamefont {Harel}},
  \bibinfo {author} {\bibfnamefont {E.}~\bibnamefont {Cohen}}, \bibinfo
  {author} {\bibfnamefont {A.}~\bibnamefont {Ron}}, \bibinfo {author}
  {\bibfnamefont {E.}~\bibnamefont {Linder}},\ and\ \bibinfo {author}
  {\bibfnamefont {L.~N.}\ \bibnamefont {Pfeiffer}},\ }\href
  {https://doi.org/10.1103/physrevlett.84.1607} {\bibfield  {journal} {\bibinfo
   {journal} {Phys. Rev. Lett.}\ }\textbf {\bibinfo {volume} {84}},\ \bibinfo
  {pages} {1607} (\bibinfo {year} {2000})}\BibitemShut {NoStop}%
\bibitem [{\citenamefont {Rapaport}\ \emph {et~al.}(2001)\citenamefont
  {Rapaport}, \citenamefont {Cohen}, \citenamefont {Ron}, \citenamefont
  {Linder},\ and\ \citenamefont {Pfeiffer}}]{Rapaport:PRB63(2001)}%
  \BibitemOpen
  \bibfield  {author} {\bibinfo {author} {\bibfnamefont {R.}~\bibnamefont
  {Rapaport}}, \bibinfo {author} {\bibfnamefont {E.}~\bibnamefont {Cohen}},
  \bibinfo {author} {\bibfnamefont {A.}~\bibnamefont {Ron}}, \bibinfo {author}
  {\bibfnamefont {E.}~\bibnamefont {Linder}},\ and\ \bibinfo {author}
  {\bibfnamefont {L.~N.}\ \bibnamefont {Pfeiffer}},\ }\href
  {https://doi.org/10.1103/physrevb.63.235310} {\bibfield  {journal} {\bibinfo
  {journal} {Phys. Rev. B}\ }\textbf {\bibinfo {volume} {63}},\ \bibinfo
  {pages} {235310} (\bibinfo {year} {2001})}\BibitemShut {NoStop}%
\bibitem [{\citenamefont {Teran}\ \emph {et~al.}(2005)\citenamefont {Teran},
  \citenamefont {Eaves}, \citenamefont {Mansouri}, \citenamefont {Buhmann},
  \citenamefont {Maude}, \citenamefont {Potemski}, \citenamefont {Henini},\
  and\ \citenamefont {Hill}}]{Teran2005}%
  \BibitemOpen
  \bibfield  {author} {\bibinfo {author} {\bibfnamefont {F.~J.}\ \bibnamefont
  {Teran}}, \bibinfo {author} {\bibfnamefont {L.}~\bibnamefont {Eaves}},
  \bibinfo {author} {\bibfnamefont {L.}~\bibnamefont {Mansouri}}, \bibinfo
  {author} {\bibfnamefont {H.}~\bibnamefont {Buhmann}}, \bibinfo {author}
  {\bibfnamefont {D.~K.}\ \bibnamefont {Maude}}, \bibinfo {author}
  {\bibfnamefont {M.}~\bibnamefont {Potemski}}, \bibinfo {author}
  {\bibfnamefont {M.}~\bibnamefont {Henini}},\ and\ \bibinfo {author}
  {\bibfnamefont {G.}~\bibnamefont {Hill}},\ }\href
  {https://doi.org/10.1103/PhysRevB.71.161309} {\bibfield  {journal} {\bibinfo
  {journal} {Phys. Rev. B}\ }\textbf {\bibinfo {volume} {71}},\ \bibinfo
  {pages} {161309} (\bibinfo {year} {2005})}\BibitemShut {NoStop}%
\bibitem [{\citenamefont {Ganchev}\ \emph {et~al.}(2015)\citenamefont
  {Ganchev}, \citenamefont {Drummond}, \citenamefont {Aleiner},\ and\
  \citenamefont {Fal'ko}}]{Ganchev:PRL114(2015)}%
  \BibitemOpen
  \bibfield  {author} {\bibinfo {author} {\bibfnamefont {B.}~\bibnamefont
  {Ganchev}}, \bibinfo {author} {\bibfnamefont {N.}~\bibnamefont {Drummond}},
  \bibinfo {author} {\bibfnamefont {I.}~\bibnamefont {Aleiner}},\ and\ \bibinfo
  {author} {\bibfnamefont {V.}~\bibnamefont {Fal'ko}},\ }\href
  {https://doi.org/10.1103/physrevlett.114.107401} {\bibfield  {journal}
  {\bibinfo  {journal} {Phys. Rev. Lett.}\ }\textbf {\bibinfo {volume} {114}},\
  \bibinfo {pages} {107401} (\bibinfo {year} {2015})}\BibitemShut {NoStop}%
\bibitem [{\citenamefont {Efimkin}\ and\ \citenamefont
  {MacDonald}(2017)}]{Efimkin:PRB95(2017)}%
  \BibitemOpen
  \bibfield  {author} {\bibinfo {author} {\bibfnamefont {D.~K.}\ \bibnamefont
  {Efimkin}}\ and\ \bibinfo {author} {\bibfnamefont {A.~H.}\ \bibnamefont
  {MacDonald}},\ }\href {https://doi.org/10.1103/physrevb.95.035417} {\bibfield
   {journal} {\bibinfo  {journal} {Phys. Rev. B}\ }\textbf {\bibinfo {volume}
  {95}},\ \bibinfo {pages} {035417} (\bibinfo {year} {2017})}\BibitemShut
  {NoStop}%
\bibitem [{\citenamefont {Courtade}\ \emph {et~al.}(2017)\citenamefont
  {Courtade}, \citenamefont {Semina}, \citenamefont {Manca}, \citenamefont
  {Glazov}, \citenamefont {Robert}, \citenamefont {Cadiz}, \citenamefont
  {Wang}, \citenamefont {Taniguchi}, \citenamefont {Watanabe}, \citenamefont
  {Pierre}, \citenamefont {Escoffier}, \citenamefont {Ivchenko}, \citenamefont
  {Renucci}, \citenamefont {Marie}, \citenamefont {Amand},\ and\ \citenamefont
  {Urbaszek}}]{Courtade:PRB96(2017)}%
  \BibitemOpen
  \bibfield  {author} {\bibinfo {author} {\bibfnamefont {E.}~\bibnamefont
  {Courtade}}, \bibinfo {author} {\bibfnamefont {M.}~\bibnamefont {Semina}},
  \bibinfo {author} {\bibfnamefont {M.}~\bibnamefont {Manca}}, \bibinfo
  {author} {\bibfnamefont {M.~M.}\ \bibnamefont {Glazov}}, \bibinfo {author}
  {\bibfnamefont {C.}~\bibnamefont {Robert}}, \bibinfo {author} {\bibfnamefont
  {F.}~\bibnamefont {Cadiz}}, \bibinfo {author} {\bibfnamefont
  {G.}~\bibnamefont {Wang}}, \bibinfo {author} {\bibfnamefont {T.}~\bibnamefont
  {Taniguchi}}, \bibinfo {author} {\bibfnamefont {K.}~\bibnamefont {Watanabe}},
  \bibinfo {author} {\bibfnamefont {M.}~\bibnamefont {Pierre}}, \bibinfo
  {author} {\bibfnamefont {W.}~\bibnamefont {Escoffier}}, \bibinfo {author}
  {\bibfnamefont {E.~L.}\ \bibnamefont {Ivchenko}}, \bibinfo {author}
  {\bibfnamefont {P.}~\bibnamefont {Renucci}}, \bibinfo {author} {\bibfnamefont
  {X.}~\bibnamefont {Marie}}, \bibinfo {author} {\bibfnamefont
  {T.}~\bibnamefont {Amand}},\ and\ \bibinfo {author} {\bibfnamefont
  {B.}~\bibnamefont {Urbaszek}},\ }\href
  {https://doi.org/10.1103/PhysRevB.96.085302} {\bibfield  {journal} {\bibinfo
  {journal} {Phys. Rev. B}\ }\textbf {\bibinfo {volume} {96}},\ \bibinfo
  {pages} {085302} (\bibinfo {year} {2017})}\BibitemShut {NoStop}%
\bibitem [{\citenamefont {Zhumagulov}\ \emph
  {et~al.}(2020{\natexlab{a}})\citenamefont {Zhumagulov}, \citenamefont
  {Vagov}, \citenamefont {Senkevich}, \citenamefont {Gulevich},\ and\
  \citenamefont {Perebeinos}}]{ZhumagulovPRB2020}%
  \BibitemOpen
  \bibfield  {author} {\bibinfo {author} {\bibfnamefont {Y.~V.}\ \bibnamefont
  {Zhumagulov}}, \bibinfo {author} {\bibfnamefont {A.}~\bibnamefont {Vagov}},
  \bibinfo {author} {\bibfnamefont {N.~Y.}\ \bibnamefont {Senkevich}}, \bibinfo
  {author} {\bibfnamefont {D.~R.}\ \bibnamefont {Gulevich}},\ and\ \bibinfo
  {author} {\bibfnamefont {V.}~\bibnamefont {Perebeinos}},\ }\href
  {https://doi.org/10.1103/PhysRevB.101.245433} {\bibfield  {journal} {\bibinfo
   {journal} {Phys. Rev. B}\ }\textbf {\bibinfo {volume} {101}},\ \bibinfo
  {pages} {245433} (\bibinfo {year} {2020}{\natexlab{a}})}\BibitemShut
  {NoStop}%
\bibitem [{\citenamefont {Shiau}\ \emph {et~al.}(2017)\citenamefont {Shiau},
  \citenamefont {Combescot},\ and\ \citenamefont {Chang}}]{Shiau_2017}%
  \BibitemOpen
  \bibfield  {author} {\bibinfo {author} {\bibfnamefont {S.-Y.}\ \bibnamefont
  {Shiau}}, \bibinfo {author} {\bibfnamefont {M.}~\bibnamefont {Combescot}},\
  and\ \bibinfo {author} {\bibfnamefont {Y.-C.}\ \bibnamefont {Chang}},\ }\href
  {https://doi.org/10.1209/0295-5075/117/57001} {\bibfield  {journal} {\bibinfo
   {journal} {{EPL} (Europhysics Letters)}\ }\textbf {\bibinfo {volume}
  {117}},\ \bibinfo {pages} {57001} (\bibinfo {year} {2017})}\BibitemShut
  {NoStop}%
\bibitem [{\citenamefont {Ravets}\ \emph {et~al.}(2018)\citenamefont {Ravets},
  \citenamefont {Knüppel}, \citenamefont {Faelt}, \citenamefont {Cotlet},
  \citenamefont {Kroner}, \citenamefont {Wegscheider},\ and\ \citenamefont
  {Imamoglu}}]{Ravets:PRL120(2018)}%
  \BibitemOpen
  \bibfield  {author} {\bibinfo {author} {\bibfnamefont {S.}~\bibnamefont
  {Ravets}}, \bibinfo {author} {\bibfnamefont {P.}~\bibnamefont {Knüppel}},
  \bibinfo {author} {\bibfnamefont {S.}~\bibnamefont {Faelt}}, \bibinfo
  {author} {\bibfnamefont {O.}~\bibnamefont {Cotlet}}, \bibinfo {author}
  {\bibfnamefont {M.}~\bibnamefont {Kroner}}, \bibinfo {author} {\bibfnamefont
  {W.}~\bibnamefont {Wegscheider}},\ and\ \bibinfo {author} {\bibfnamefont
  {A.}~\bibnamefont {Imamoglu}},\ }\href
  {https://doi.org/10.1103/physrevlett.120.057401} {\bibfield  {journal}
  {\bibinfo  {journal} {Phys. Rev. Lett.}\ }\textbf {\bibinfo {volume} {120}},\
  \bibinfo {pages} {057401} (\bibinfo {year} {2018})}\BibitemShut {NoStop}%
\bibitem [{\citenamefont {Chang}\ \emph {et~al.}(2018)\citenamefont {Chang},
  \citenamefont {Shiau},\ and\ \citenamefont {Combescot}}]{Chang:PRB98(2018)}%
  \BibitemOpen
  \bibfield  {author} {\bibinfo {author} {\bibfnamefont {Y.-C.}\ \bibnamefont
  {Chang}}, \bibinfo {author} {\bibfnamefont {S.-Y.}\ \bibnamefont {Shiau}},\
  and\ \bibinfo {author} {\bibfnamefont {M.}~\bibnamefont {Combescot}},\ }\href
  {https://doi.org/10.1103/physrevb.98.235203} {\bibfield  {journal} {\bibinfo
  {journal} {Phys. Rev. B}\ }\textbf {\bibinfo {volume} {98}},\ \bibinfo
  {pages} {235203} (\bibinfo {year} {2018})}\BibitemShut {NoStop}%
\bibitem [{\citenamefont {Levinsen}\ \emph {et~al.}(2019)\citenamefont
  {Levinsen}, \citenamefont {Li},\ and\ \citenamefont
  {Parish}}]{Levinsen:PRR1(2019)}%
  \BibitemOpen
  \bibfield  {author} {\bibinfo {author} {\bibfnamefont {J.}~\bibnamefont
  {Levinsen}}, \bibinfo {author} {\bibfnamefont {G.}~\bibnamefont {Li}},\ and\
  \bibinfo {author} {\bibfnamefont {M.~M.}\ \bibnamefont {Parish}},\ }\href
  {https://doi.org/10.1103/physrevresearch.1.033120} {\bibfield  {journal}
  {\bibinfo  {journal} {Phys. Rev. Research}\ }\textbf {\bibinfo {volume}
  {1}},\ \bibinfo {pages} {033120} (\bibinfo {year} {2019})}\BibitemShut
  {NoStop}%
\bibitem [{\citenamefont {Rana}\ \emph {et~al.}(2020)\citenamefont {Rana},
  \citenamefont {Koksal},\ and\ \citenamefont {Manolatou}}]{Rana:PRB102(2020)}%
  \BibitemOpen
  \bibfield  {author} {\bibinfo {author} {\bibfnamefont {F.}~\bibnamefont
  {Rana}}, \bibinfo {author} {\bibfnamefont {O.}~\bibnamefont {Koksal}},\ and\
  \bibinfo {author} {\bibfnamefont {C.}~\bibnamefont {Manolatou}},\ }\href
  {https://doi.org/10.1103/physrevb.102.085304} {\bibfield  {journal} {\bibinfo
   {journal} {Phys. Rev. B}\ }\textbf {\bibinfo {volume} {102}},\ \bibinfo
  {pages} {085304} (\bibinfo {year} {2020})}\BibitemShut {NoStop}%
\bibitem [{\citenamefont {Rana}\ \emph {et~al.}(2021)\citenamefont {Rana},
  \citenamefont {Koksal}, \citenamefont {Jung}, \citenamefont {Shvets},
  \citenamefont {Vamivakas},\ and\ \citenamefont
  {Manolatou}}]{Rana:PRL126(2021)}%
  \BibitemOpen
  \bibfield  {author} {\bibinfo {author} {\bibfnamefont {F.}~\bibnamefont
  {Rana}}, \bibinfo {author} {\bibfnamefont {O.}~\bibnamefont {Koksal}},
  \bibinfo {author} {\bibfnamefont {M.}~\bibnamefont {Jung}}, \bibinfo {author}
  {\bibfnamefont {G.}~\bibnamefont {Shvets}}, \bibinfo {author} {\bibfnamefont
  {A.}~\bibnamefont {Vamivakas}},\ and\ \bibinfo {author} {\bibfnamefont
  {C.}~\bibnamefont {Manolatou}},\ }\href
  {https://doi.org/10.1103/physrevlett.126.127402} {\bibfield  {journal}
  {\bibinfo  {journal} {Phys. Rev. Lett.}\ }\textbf {\bibinfo {volume} {126}},\
  \bibinfo {pages} {127402} (\bibinfo {year} {2021})}\BibitemShut {NoStop}%
\bibitem [{\citenamefont {Koksal}\ \emph {et~al.}(2021)\citenamefont {Koksal},
  \citenamefont {Jung}, \citenamefont {Manolatou}, \citenamefont {Vamivakas},
  \citenamefont {Shvets},\ and\ \citenamefont {Rana}}]{Koksal:PRR3(2021)}%
  \BibitemOpen
  \bibfield  {author} {\bibinfo {author} {\bibfnamefont {O.}~\bibnamefont
  {Koksal}}, \bibinfo {author} {\bibfnamefont {M.}~\bibnamefont {Jung}},
  \bibinfo {author} {\bibfnamefont {C.}~\bibnamefont {Manolatou}}, \bibinfo
  {author} {\bibfnamefont {A.~N.}\ \bibnamefont {Vamivakas}}, \bibinfo {author}
  {\bibfnamefont {G.}~\bibnamefont {Shvets}},\ and\ \bibinfo {author}
  {\bibfnamefont {F.}~\bibnamefont {Rana}},\ }\href
  {https://doi.org/10.1103/physrevresearch.3.033064} {\bibfield  {journal}
  {\bibinfo  {journal} {Phys. Rev. Research}\ }\textbf {\bibinfo {volume}
  {3}},\ \bibinfo {pages} {033064} (\bibinfo {year} {2021})}\BibitemShut
  {NoStop}%
\bibitem [{\citenamefont {Kyriienko}\ \emph {et~al.}(2020)\citenamefont
  {Kyriienko}, \citenamefont {Krizhanovskii},\ and\ \citenamefont
  {Shelykh}}]{Kyriienko:PRL125(2020)}%
  \BibitemOpen
  \bibfield  {author} {\bibinfo {author} {\bibfnamefont {O.}~\bibnamefont
  {Kyriienko}}, \bibinfo {author} {\bibfnamefont {D.}~\bibnamefont
  {Krizhanovskii}},\ and\ \bibinfo {author} {\bibfnamefont {I.}~\bibnamefont
  {Shelykh}},\ }\href {https://doi.org/10.1103/physrevlett.125.197402}
  {\bibfield  {journal} {\bibinfo  {journal} {Phys. Rev. Lett.}\ }\textbf
  {\bibinfo {volume} {125}},\ \bibinfo {pages} {197402} (\bibinfo {year}
  {2020})}\BibitemShut {NoStop}%
\bibitem [{\citenamefont {Zhumagulov}\ \emph {et~al.}(2022)\citenamefont
  {Zhumagulov}, \citenamefont {Chiavazzo}, \citenamefont {Perebeinos},
  \citenamefont {Shelykh},\ and\ \citenamefont
  {Kyriienko}}]{Zhumagulov:NPJ2022}%
  \BibitemOpen
  \bibfield  {author} {\bibinfo {author} {\bibfnamefont {Y.}~\bibnamefont
  {Zhumagulov}}, \bibinfo {author} {\bibfnamefont {D.}~\bibnamefont
  {Chiavazzo}, \bibfnamefont {S.and~Gulevich}}, \bibinfo {author}
  {\bibfnamefont {V.}~\bibnamefont {Perebeinos}}, \bibinfo {author}
  {\bibfnamefont {I.}~\bibnamefont {Shelykh}},\ and\ \bibinfo {author}
  {\bibfnamefont {O.}~\bibnamefont {Kyriienko}},\ }\href
  {https://doi.org/https://doi.org/10.1038/s41524-022-00775-x} {\bibfield
  {journal} {\bibinfo  {journal} {npj Comput. Mater.}\ }\textbf {\bibinfo
  {volume} {8}},\ \bibinfo {pages} {92} (\bibinfo {year} {2022})}\BibitemShut
  {NoStop}%
\bibitem [{\citenamefont {Bastarrachea-Magnani}\ \emph
  {et~al.}(2021)\citenamefont {Bastarrachea-Magnani}, \citenamefont
  {Camacho-Guardian},\ and\ \citenamefont
  {Bruun}}]{BastarracheaMagnani:PRL126(2021)}%
  \BibitemOpen
  \bibfield  {author} {\bibinfo {author} {\bibfnamefont {M.~A.}\ \bibnamefont
  {Bastarrachea-Magnani}}, \bibinfo {author} {\bibfnamefont {A.}~\bibnamefont
  {Camacho-Guardian}},\ and\ \bibinfo {author} {\bibfnamefont {G.~M.}\
  \bibnamefont {Bruun}},\ }\href
  {https://doi.org/10.1103/physrevlett.126.127405} {\bibfield  {journal}
  {\bibinfo  {journal} {Phys. Rev. Lett.}\ }\textbf {\bibinfo {volume} {126}},\
  \bibinfo {pages} {127405} (\bibinfo {year} {2021})}\BibitemShut {NoStop}%
\bibitem [{\citenamefont {Li}\ \emph {et~al.}(2021{\natexlab{a}})\citenamefont
  {Li}, \citenamefont {Bleu}, \citenamefont {Parish},\ and\ \citenamefont
  {Levinsen}}]{LiBleuPRL2021}%
  \BibitemOpen
  \bibfield  {author} {\bibinfo {author} {\bibfnamefont {G.}~\bibnamefont
  {Li}}, \bibinfo {author} {\bibfnamefont {O.}~\bibnamefont {Bleu}}, \bibinfo
  {author} {\bibfnamefont {M.~M.}\ \bibnamefont {Parish}},\ and\ \bibinfo
  {author} {\bibfnamefont {J.}~\bibnamefont {Levinsen}},\ }\href
  {https://doi.org/10.1103/PhysRevLett.126.197401} {\bibfield  {journal}
  {\bibinfo  {journal} {Phys. Rev. Lett.}\ }\textbf {\bibinfo {volume} {126}},\
  \bibinfo {pages} {197401} (\bibinfo {year} {2021}{\natexlab{a}})}\BibitemShut
  {NoStop}%
\bibitem [{\citenamefont {Li}\ \emph {et~al.}(2021{\natexlab{b}})\citenamefont
  {Li}, \citenamefont {Bleu}, \citenamefont {Levinsen},\ and\ \citenamefont
  {Parish}}]{LiBleuPRB2021}%
  \BibitemOpen
  \bibfield  {author} {\bibinfo {author} {\bibfnamefont {G.}~\bibnamefont
  {Li}}, \bibinfo {author} {\bibfnamefont {O.}~\bibnamefont {Bleu}}, \bibinfo
  {author} {\bibfnamefont {J.}~\bibnamefont {Levinsen}},\ and\ \bibinfo
  {author} {\bibfnamefont {M.~M.}\ \bibnamefont {Parish}},\ }\href
  {https://doi.org/10.1103/PhysRevB.103.195307} {\bibfield  {journal} {\bibinfo
   {journal} {Phys. Rev. B}\ }\textbf {\bibinfo {volume} {103}},\ \bibinfo
  {pages} {195307} (\bibinfo {year} {2021}{\natexlab{b}})}\BibitemShut
  {NoStop}%
\bibitem [{\citenamefont {Efimkin}\ \emph
  {et~al.}(2021{\natexlab{a}})\citenamefont {Efimkin}, \citenamefont {Laird},
  \citenamefont {Levinsen}, \citenamefont {Parish},\ and\ \citenamefont
  {MacDonald}}]{EfimkinPRB2021}%
  \BibitemOpen
  \bibfield  {author} {\bibinfo {author} {\bibfnamefont {D.~K.}\ \bibnamefont
  {Efimkin}}, \bibinfo {author} {\bibfnamefont {E.~K.}\ \bibnamefont {Laird}},
  \bibinfo {author} {\bibfnamefont {J.}~\bibnamefont {Levinsen}}, \bibinfo
  {author} {\bibfnamefont {M.~M.}\ \bibnamefont {Parish}},\ and\ \bibinfo
  {author} {\bibfnamefont {A.~H.}\ \bibnamefont {MacDonald}},\ }\href
  {https://doi.org/10.1103/PhysRevB.103.075417} {\bibfield  {journal} {\bibinfo
   {journal} {Phys. Rev. B}\ }\textbf {\bibinfo {volume} {103}},\ \bibinfo
  {pages} {075417} (\bibinfo {year} {2021}{\natexlab{a}})}\BibitemShut
  {NoStop}%
\bibitem [{\citenamefont {Tiene}\ \emph {et~al.}(2022)\citenamefont {Tiene},
  \citenamefont {Levinsen}, \citenamefont {Keeling}, \citenamefont {Parish},\
  and\ \citenamefont {Marchetti}}]{Tiene2022}%
  \BibitemOpen
  \bibfield  {author} {\bibinfo {author} {\bibfnamefont {A.}~\bibnamefont
  {Tiene}}, \bibinfo {author} {\bibfnamefont {J.}~\bibnamefont {Levinsen}},
  \bibinfo {author} {\bibfnamefont {J.}~\bibnamefont {Keeling}}, \bibinfo
  {author} {\bibfnamefont {M.~M.}\ \bibnamefont {Parish}},\ and\ \bibinfo
  {author} {\bibfnamefont {F.~M.}\ \bibnamefont {Marchetti}},\ }\href
  {https://doi.org/10.1103/PhysRevB.105.125404} {\bibfield  {journal} {\bibinfo
   {journal} {Phys. Rev. B}\ }\textbf {\bibinfo {volume} {105}},\ \bibinfo
  {pages} {125404} (\bibinfo {year} {2022})}\BibitemShut {NoStop}%
\bibitem [{\citenamefont {Denning}\ \emph
  {et~al.}(2022{\natexlab{a}})\citenamefont {Denning}, \citenamefont {Wubs},
  \citenamefont {Stenger}, \citenamefont {M{\o}rk},\ and\ \citenamefont
  {Kristensen}}]{Denning:PRB105(2022)}%
  \BibitemOpen
  \bibfield  {author} {\bibinfo {author} {\bibfnamefont {E.~V.}\ \bibnamefont
  {Denning}}, \bibinfo {author} {\bibfnamefont {M.}~\bibnamefont {Wubs}},
  \bibinfo {author} {\bibfnamefont {N.}~\bibnamefont {Stenger}}, \bibinfo
  {author} {\bibfnamefont {J.}~\bibnamefont {M{\o}rk}},\ and\ \bibinfo {author}
  {\bibfnamefont {P.~T.}\ \bibnamefont {Kristensen}},\ }\href
  {https://doi.org/10.1103/physrevb.105.085306} {\bibfield  {journal} {\bibinfo
   {journal} {Phys. Rev. B}\ }\textbf {\bibinfo {volume} {105}},\ \bibinfo
  {pages} {085306} (\bibinfo {year} {2022}{\natexlab{a}})}\BibitemShut
  {NoStop}%
\bibitem [{\citenamefont {Denning}\ \emph
  {et~al.}(2022{\natexlab{b}})\citenamefont {Denning}, \citenamefont {Wubs},
  \citenamefont {Stenger}, \citenamefont {M\o{}rk},\ and\ \citenamefont
  {Kristensen}}]{Denning:Arxiv(2021)}%
  \BibitemOpen
  \bibfield  {author} {\bibinfo {author} {\bibfnamefont {E.~V.}\ \bibnamefont
  {Denning}}, \bibinfo {author} {\bibfnamefont {M.}~\bibnamefont {Wubs}},
  \bibinfo {author} {\bibfnamefont {N.}~\bibnamefont {Stenger}}, \bibinfo
  {author} {\bibfnamefont {J.}~\bibnamefont {M\o{}rk}},\ and\ \bibinfo {author}
  {\bibfnamefont {P.~T.}\ \bibnamefont {Kristensen}},\ }\href
  {https://doi.org/10.1103/PhysRevResearch.4.L012020} {\bibfield  {journal}
  {\bibinfo  {journal} {Phys. Rev. Research}\ }\textbf {\bibinfo {volume}
  {4}},\ \bibinfo {pages} {L012020} (\bibinfo {year}
  {2022}{\natexlab{b}})}\BibitemShut {NoStop}%
\bibitem [{\citenamefont {Combescot}(2017)}]{Combescot:PRX7(2017)}%
  \BibitemOpen
  \bibfield  {author} {\bibinfo {author} {\bibfnamefont {R.}~\bibnamefont
  {Combescot}},\ }\href {https://doi.org/10.1103/physrevx.7.041035} {\bibfield
  {journal} {\bibinfo  {journal} {Phys. Rev. X}\ }\textbf {\bibinfo {volume}
  {7}},\ \bibinfo {pages} {041035} (\bibinfo {year} {2017})}\BibitemShut
  {NoStop}%
\bibitem [{\citenamefont {Combescot}\ and\ \citenamefont
  {Betbeder-Matibet}(2011)}]{Combescot:EurPhysJB79(2011)}%
  \BibitemOpen
  \bibfield  {author} {\bibinfo {author} {\bibfnamefont {M.}~\bibnamefont
  {Combescot}}\ and\ \bibinfo {author} {\bibfnamefont {O.}~\bibnamefont
  {Betbeder-Matibet}},\ }\href {https://doi.org/10.1140/epjb/e2010-10804-6}
  {\bibfield  {journal} {\bibinfo  {journal} {Eur. Phys. J. B}\ }\textbf
  {\bibinfo {volume} {79}},\ \bibinfo {pages} {401} (\bibinfo {year}
  {2011})}\BibitemShut {NoStop}%
\bibitem [{\citenamefont {Shiau}\ \emph {et~al.}(2012)\citenamefont {Shiau},
  \citenamefont {Combescot},\ and\ \citenamefont {Chang}}]{Shiau:PRB86(2012)}%
  \BibitemOpen
  \bibfield  {author} {\bibinfo {author} {\bibfnamefont {S.-Y.}\ \bibnamefont
  {Shiau}}, \bibinfo {author} {\bibfnamefont {M.}~\bibnamefont {Combescot}},\
  and\ \bibinfo {author} {\bibfnamefont {Y.-C.}\ \bibnamefont {Chang}},\ }\href
  {https://doi.org/10.1103/physrevb.86.115210} {\bibfield  {journal} {\bibinfo
  {journal} {Phys. Rev. B}\ }\textbf {\bibinfo {volume} {86}},\ \bibinfo
  {pages} {115210} (\bibinfo {year} {2012})}\BibitemShut {NoStop}%
\bibitem [{\citenamefont {Song}\ \emph {et~al.}(2022)\citenamefont {Song},
  \citenamefont {Chiavazzo}, \citenamefont {Shelykh},\ and\ \citenamefont
  {Kyriienko}}]{Song2022b}%
  \BibitemOpen
  \bibfield  {author} {\bibinfo {author} {\bibfnamefont {K.~W.}\ \bibnamefont
  {Song}}, \bibinfo {author} {\bibfnamefont {S.}~\bibnamefont {Chiavazzo}},
  \bibinfo {author} {\bibfnamefont {I.~A.}\ \bibnamefont {Shelykh}},\ and\
  \bibinfo {author} {\bibfnamefont {O.}~\bibnamefont {Kyriienko}},\ }\href@noop
  {} {\bibinfo {title} {Theory for coulomb scattering of trions in 2d
  materials}} (\bibinfo {year} {2022}),\ \Eprint
  {https://arxiv.org/abs/submitted} {submitted} \BibitemShut {NoStop}%
\bibitem [{\citenamefont {{Keldysh}}(1979)}]{Keldysh:JETP29-1979}%
  \BibitemOpen
  \bibfield  {author} {\bibinfo {author} {\bibfnamefont {L.~V.}\ \bibnamefont
  {{Keldysh}}},\ }\href@noop {} {\bibfield  {journal} {\bibinfo  {journal}
  {Soviet Journal of Experimental and Theoretical Physics Letters}\ }\textbf
  {\bibinfo {volume} {29}},\ \bibinfo {pages} {658} (\bibinfo {year}
  {1979})}\BibitemShut {NoStop}%
\bibitem [{\citenamefont {Rytova}(1967)}]{Rytova:MUPhys3-1967}%
  \BibitemOpen
  \bibfield  {author} {\bibinfo {author} {\bibfnamefont {N.~S.}\ \bibnamefont
  {Rytova}},\ }\href@noop {} {\bibfield  {journal} {\bibinfo  {journal} {Moscow
  Univ. Phys. Bull.}\ }\textbf {\bibinfo {volume} {3}},\ \bibinfo {pages} {18}
  (\bibinfo {year} {1967})}\BibitemShut {NoStop}%
\bibitem [{\citenamefont {Chevy}(2006)}]{Chevy:PRA74(2006)}%
  \BibitemOpen
  \bibfield  {author} {\bibinfo {author} {\bibfnamefont {F.}~\bibnamefont
  {Chevy}},\ }\href {https://doi.org/10.1103/physreva.74.063628} {\bibfield
  {journal} {\bibinfo  {journal} {Physical Review A}\ }\textbf {\bibinfo
  {volume} {74}},\ \bibinfo {pages} {063628} (\bibinfo {year}
  {2006})}\BibitemShut {NoStop}%
\bibitem [{\citenamefont {Combescot}\ and\ \citenamefont
  {Giraud}(2008)}]{Combescot:PRL101(2008)}%
  \BibitemOpen
  \bibfield  {author} {\bibinfo {author} {\bibfnamefont {R.}~\bibnamefont
  {Combescot}}\ and\ \bibinfo {author} {\bibfnamefont {S.}~\bibnamefont
  {Giraud}},\ }\href {https://doi.org/10.1103/physrevlett.101.050404}
  {\bibfield  {journal} {\bibinfo  {journal} {Phys. Rev. Lett.}\ }\textbf
  {\bibinfo {volume} {101}},\ \bibinfo {pages} {050404} (\bibinfo {year}
  {2008})}\BibitemShut {NoStop}%
\bibitem [{\citenamefont {Efimkin}\ \emph
  {et~al.}(2021{\natexlab{b}})\citenamefont {Efimkin}, \citenamefont {Laird},
  \citenamefont {Levinsen}, \citenamefont {Parish},\ and\ \citenamefont
  {MacDonald}}]{Efimkin:PRB103(2021)}%
  \BibitemOpen
  \bibfield  {author} {\bibinfo {author} {\bibfnamefont {D.~K.}\ \bibnamefont
  {Efimkin}}, \bibinfo {author} {\bibfnamefont {E.~K.}\ \bibnamefont {Laird}},
  \bibinfo {author} {\bibfnamefont {J.}~\bibnamefont {Levinsen}}, \bibinfo
  {author} {\bibfnamefont {M.~M.}\ \bibnamefont {Parish}},\ and\ \bibinfo
  {author} {\bibfnamefont {A.~H.}\ \bibnamefont {MacDonald}},\ }\href
  {https://doi.org/10.1103/physrevb.103.075417} {\bibfield  {journal} {\bibinfo
   {journal} {Phys. Rev. B}\ }\textbf {\bibinfo {volume} {103}},\ \bibinfo
  {pages} {075417} (\bibinfo {year} {2021}{\natexlab{b}})}\BibitemShut
  {NoStop}%
\bibitem [{\citenamefont {Li}\ \emph {et~al.}(2021{\natexlab{c}})\citenamefont
  {Li}, \citenamefont {Bleu}, \citenamefont {Parish},\ and\ \citenamefont
  {Levinsen}}]{Li:PRL126(2021)}%
  \BibitemOpen
  \bibfield  {author} {\bibinfo {author} {\bibfnamefont {G.}~\bibnamefont
  {Li}}, \bibinfo {author} {\bibfnamefont {O.}~\bibnamefont {Bleu}}, \bibinfo
  {author} {\bibfnamefont {M.~M.}\ \bibnamefont {Parish}},\ and\ \bibinfo
  {author} {\bibfnamefont {J.}~\bibnamefont {Levinsen}},\ }\href
  {https://doi.org/10.1103/physrevlett.126.197401} {\bibfield  {journal}
  {\bibinfo  {journal} {Phys. Rev. Lett.}\ }\textbf {\bibinfo {volume} {126}},\
  \bibinfo {pages} {197401} (\bibinfo {year} {2021}{\natexlab{c}})}\BibitemShut
  {NoStop}%
\bibitem [{\citenamefont {Li}\ \emph {et~al.}(2021{\natexlab{d}})\citenamefont
  {Li}, \citenamefont {Bleu}, \citenamefont {Levinsen},\ and\ \citenamefont
  {Parish}}]{Li:PRB103(2021)}%
  \BibitemOpen
  \bibfield  {author} {\bibinfo {author} {\bibfnamefont {G.}~\bibnamefont
  {Li}}, \bibinfo {author} {\bibfnamefont {O.}~\bibnamefont {Bleu}}, \bibinfo
  {author} {\bibfnamefont {J.}~\bibnamefont {Levinsen}},\ and\ \bibinfo
  {author} {\bibfnamefont {M.~M.}\ \bibnamefont {Parish}},\ }\href
  {https://doi.org/10.1103/physrevb.103.195307} {\bibfield  {journal} {\bibinfo
   {journal} {Phys. Rev. B}\ }\textbf {\bibinfo {volume} {103}},\ \bibinfo
  {pages} {195307} (\bibinfo {year} {2021}{\natexlab{d}})}\BibitemShut
  {NoStop}%
\bibitem [{\citenamefont {Glazov}(2020)}]{Glazov:JChemPhys153(2020)}%
  \BibitemOpen
  \bibfield  {author} {\bibinfo {author} {\bibfnamefont {M.~M.}\ \bibnamefont
  {Glazov}},\ }\href {https://doi.org/10.1063/5.0012475} {\bibfield  {journal}
  {\bibinfo  {journal} {J. Chem. Phys.}\ }\textbf {\bibinfo {volume} {153}},\
  \bibinfo {pages} {034703} (\bibinfo {year} {2020})}\BibitemShut {NoStop}%
\bibitem [{\citenamefont {Combescot}\ \emph {et~al.}(2008)\citenamefont
  {Combescot}, \citenamefont {Betbeder-Matibet},\ and\ \citenamefont
  {Dubi}}]{Combescot:PhysRep463(2008)}%
  \BibitemOpen
  \bibfield  {author} {\bibinfo {author} {\bibfnamefont {M.}~\bibnamefont
  {Combescot}}, \bibinfo {author} {\bibfnamefont {O.}~\bibnamefont
  {Betbeder-Matibet}},\ and\ \bibinfo {author} {\bibfnamefont {F.}~\bibnamefont
  {Dubi}},\ }\href {https://doi.org/10.1016/j.physrep.2007.11.003} {\bibfield
  {journal} {\bibinfo  {journal} {Physics Reports}\ }\textbf {\bibinfo {volume}
  {463}},\ \bibinfo {pages} {215} (\bibinfo {year} {2008})}\BibitemShut
  {NoStop}%
\bibitem [{\citenamefont {Deilmann}\ \emph {et~al.}(2016)\citenamefont
  {Deilmann}, \citenamefont {Drüppel},\ and\ \citenamefont
  {Rohlfing}}]{Deilmann:PRL116(2016)}%
  \BibitemOpen
  \bibfield  {author} {\bibinfo {author} {\bibfnamefont {T.}~\bibnamefont
  {Deilmann}}, \bibinfo {author} {\bibfnamefont {M.}~\bibnamefont {Drüppel}},\
  and\ \bibinfo {author} {\bibfnamefont {M.}~\bibnamefont {Rohlfing}},\ }\href
  {https://doi.org/10.1103/physrevlett.116.196804} {\bibfield  {journal}
  {\bibinfo  {journal} {Phys. Rev. Lett.}\ }\textbf {\bibinfo {volume} {116}},\
  \bibinfo {pages} {196804} (\bibinfo {year} {2016})}\BibitemShut {NoStop}%
\bibitem [{\citenamefont {Drüppel}\ \emph {et~al.}(2017)\citenamefont
  {Drüppel}, \citenamefont {Deilmann}, \citenamefont {Krüger},\ and\
  \citenamefont {Rohlfing}}]{Drueppel:NatComm8(2017)}%
  \BibitemOpen
  \bibfield  {author} {\bibinfo {author} {\bibfnamefont {M.}~\bibnamefont
  {Drüppel}}, \bibinfo {author} {\bibfnamefont {T.}~\bibnamefont {Deilmann}},
  \bibinfo {author} {\bibfnamefont {P.}~\bibnamefont {Krüger}},\ and\ \bibinfo
  {author} {\bibfnamefont {M.}~\bibnamefont {Rohlfing}},\ }\bibfield  {journal}
  {\bibinfo  {journal} {Nat. Commun.}\ }\textbf {\bibinfo {volume} {8}},\ \href
  {https://doi.org/10.1038/s41467-017-02286-6} {10.1038/s41467-017-02286-6}
  (\bibinfo {year} {2017})\BibitemShut {NoStop}%
\bibitem [{\citenamefont {Torche}\ and\ \citenamefont
  {Bester}(2019)}]{Torche:PRB100(2019)}%
  \BibitemOpen
  \bibfield  {author} {\bibinfo {author} {\bibfnamefont {A.}~\bibnamefont
  {Torche}}\ and\ \bibinfo {author} {\bibfnamefont {G.}~\bibnamefont
  {Bester}},\ }\href {https://doi.org/10.1103/physrevb.100.201403} {\bibfield
  {journal} {\bibinfo  {journal} {Phys. Rev. B}\ }\textbf {\bibinfo {volume}
  {100}},\ \bibinfo {pages} {201403} (\bibinfo {year} {2019})}\BibitemShut
  {NoStop}%
\bibitem [{\citenamefont {Zhumagulov}\ \emph
  {et~al.}(2020{\natexlab{b}})\citenamefont {Zhumagulov}, \citenamefont
  {Vagov}, \citenamefont {Senkevich}, \citenamefont {Gulevich},\ and\
  \citenamefont {Perebeinos}}]{Zhumagulov:PRB101(2020)}%
  \BibitemOpen
  \bibfield  {author} {\bibinfo {author} {\bibfnamefont {Y.~V.}\ \bibnamefont
  {Zhumagulov}}, \bibinfo {author} {\bibfnamefont {A.}~\bibnamefont {Vagov}},
  \bibinfo {author} {\bibfnamefont {N.~Y.}\ \bibnamefont {Senkevich}}, \bibinfo
  {author} {\bibfnamefont {D.~R.}\ \bibnamefont {Gulevich}},\ and\ \bibinfo
  {author} {\bibfnamefont {V.}~\bibnamefont {Perebeinos}},\ }\href
  {https://doi.org/10.1103/physrevb.101.245433} {\bibfield  {journal} {\bibinfo
   {journal} {Phys. Rev. B}\ }\textbf {\bibinfo {volume} {101}},\ \bibinfo
  {pages} {245433} (\bibinfo {year} {2020}{\natexlab{b}})}\BibitemShut
  {NoStop}%
\bibitem [{\citenamefont {Zhumagulov}\ \emph
  {et~al.}(2020{\natexlab{c}})\citenamefont {Zhumagulov}, \citenamefont
  {Vagov}, \citenamefont {Gulevich}, \citenamefont {Junior},\ and\
  \citenamefont {Perebeinos}}]{Zhumagulov:JChemPhys153(2020)}%
  \BibitemOpen
  \bibfield  {author} {\bibinfo {author} {\bibfnamefont {Y.~V.}\ \bibnamefont
  {Zhumagulov}}, \bibinfo {author} {\bibfnamefont {A.}~\bibnamefont {Vagov}},
  \bibinfo {author} {\bibfnamefont {D.~R.}\ \bibnamefont {Gulevich}}, \bibinfo
  {author} {\bibfnamefont {P.~E.~F.}\ \bibnamefont {Junior}},\ and\ \bibinfo
  {author} {\bibfnamefont {V.}~\bibnamefont {Perebeinos}},\ }\href
  {https://doi.org/10.1063/5.0012971} {\bibfield  {journal} {\bibinfo
  {journal} {J. Chem. Phys.}\ }\textbf {\bibinfo {volume} {153}},\ \bibinfo
  {pages} {044132} (\bibinfo {year} {2020}{\natexlab{c}})}\BibitemShut
  {NoStop}%
\bibitem [{\citenamefont {Kidd}\ \emph {et~al.}(2016)\citenamefont {Kidd},
  \citenamefont {Zhang},\ and\ \citenamefont {Varga}}]{Kidd:PRB93(2016)}%
  \BibitemOpen
  \bibfield  {author} {\bibinfo {author} {\bibfnamefont {D.~W.}\ \bibnamefont
  {Kidd}}, \bibinfo {author} {\bibfnamefont {D.~K.}\ \bibnamefont {Zhang}},\
  and\ \bibinfo {author} {\bibfnamefont {K.}~\bibnamefont {Varga}},\ }\href
  {https://doi.org/10.1103/physrevb.93.125423} {\bibfield  {journal} {\bibinfo
  {journal} {Phys. Rev. B}\ }\textbf {\bibinfo {volume} {93}},\ \bibinfo
  {pages} {125423} (\bibinfo {year} {2016})}\BibitemShut {NoStop}%
\bibitem [{\citenamefont {Ceferino}\ \emph {et~al.}(2020)\citenamefont
  {Ceferino}, \citenamefont {Song}, \citenamefont {Magorrian}, \citenamefont
  {Z{\'{o}}lyomi},\ and\ \citenamefont {Fal'ko}}]{Ceferino:PRB101(2020)}%
  \BibitemOpen
  \bibfield  {author} {\bibinfo {author} {\bibfnamefont {A.}~\bibnamefont
  {Ceferino}}, \bibinfo {author} {\bibfnamefont {K.~W.}\ \bibnamefont {Song}},
  \bibinfo {author} {\bibfnamefont {S.~J.}\ \bibnamefont {Magorrian}}, \bibinfo
  {author} {\bibfnamefont {V.}~\bibnamefont {Z{\'{o}}lyomi}},\ and\ \bibinfo
  {author} {\bibfnamefont {V.~I.}\ \bibnamefont {Fal'ko}},\ }\href
  {https://doi.org/10.1103/physrevb.101.245432} {\bibfield  {journal} {\bibinfo
   {journal} {Phys. Rev. B}\ }\textbf {\bibinfo {volume} {101}},\ \bibinfo
  {pages} {245432} (\bibinfo {year} {2020})}\BibitemShut {NoStop}%
\bibitem [{\citenamefont {Glazov}\ \emph {et~al.}(2009)\citenamefont {Glazov},
  \citenamefont {Ouerdane}, \citenamefont {Pilozzi}, \citenamefont {Malpuech},
  \citenamefont {Kavokin},\ and\ \citenamefont {D'Andrea}}]{Glazov2009}%
  \BibitemOpen
  \bibfield  {author} {\bibinfo {author} {\bibfnamefont {M.~M.}\ \bibnamefont
  {Glazov}}, \bibinfo {author} {\bibfnamefont {H.}~\bibnamefont {Ouerdane}},
  \bibinfo {author} {\bibfnamefont {L.}~\bibnamefont {Pilozzi}}, \bibinfo
  {author} {\bibfnamefont {G.}~\bibnamefont {Malpuech}}, \bibinfo {author}
  {\bibfnamefont {A.~V.}\ \bibnamefont {Kavokin}},\ and\ \bibinfo {author}
  {\bibfnamefont {A.}~\bibnamefont {D'Andrea}},\ }\href
  {https://doi.org/10.1103/PhysRevB.80.155306} {\bibfield  {journal} {\bibinfo
  {journal} {Phys. Rev. B}\ }\textbf {\bibinfo {volume} {80}},\ \bibinfo
  {pages} {155306} (\bibinfo {year} {2009})}\BibitemShut {NoStop}%
\bibitem [{\citenamefont {Ciuti}\ \emph {et~al.}(1998)\citenamefont {Ciuti},
  \citenamefont {Savona}, \citenamefont {Piermarocchi}, \citenamefont
  {Quattropani},\ and\ \citenamefont {Schwendimann}}]{Ciuti1998}%
  \BibitemOpen
  \bibfield  {author} {\bibinfo {author} {\bibfnamefont {C.}~\bibnamefont
  {Ciuti}}, \bibinfo {author} {\bibfnamefont {V.}~\bibnamefont {Savona}},
  \bibinfo {author} {\bibfnamefont {C.}~\bibnamefont {Piermarocchi}}, \bibinfo
  {author} {\bibfnamefont {A.}~\bibnamefont {Quattropani}},\ and\ \bibinfo
  {author} {\bibfnamefont {P.}~\bibnamefont {Schwendimann}},\ }\href
  {https://doi.org/10.1103/PhysRevB.58.7926} {\bibfield  {journal} {\bibinfo
  {journal} {Phys. Rev. B}\ }\textbf {\bibinfo {volume} {58}},\ \bibinfo
  {pages} {7926} (\bibinfo {year} {1998})}\BibitemShut {NoStop}%
\bibitem [{\citenamefont {Combescot}\ and\ \citenamefont
  {Betbeder-Matibet}(2010)}]{Combescot:PRL104(2010)}%
  \BibitemOpen
  \bibfield  {author} {\bibinfo {author} {\bibfnamefont {M.}~\bibnamefont
  {Combescot}}\ and\ \bibinfo {author} {\bibfnamefont {O.}~\bibnamefont
  {Betbeder-Matibet}},\ }\href {https://doi.org/10.1103/physrevlett.104.206404}
  {\bibfield  {journal} {\bibinfo  {journal} {Phys. Rev. Lett.}\ }\textbf
  {\bibinfo {volume} {104}},\ \bibinfo {pages} {206404} (\bibinfo {year}
  {2010})}\BibitemShut {NoStop}%
\bibitem [{\citenamefont {Mahan}(2000)}]{Mahan:Many-Particle(2000)}%
  \BibitemOpen
  \bibfield  {author} {\bibinfo {author} {\bibfnamefont {G.~D.}\ \bibnamefont
  {Mahan}},\ }\href@noop {} {\emph {\bibinfo {title} {Many-Particle
  Physics}}},\ \bibinfo {edition} {3rd}\ ed.,\ Physics of Solids and Liquids\
  (\bibinfo  {publisher} {Springer},\ \bibinfo {year} {2000})\BibitemShut
  {NoStop}%
\bibitem [{\citenamefont {Shahnazaryan}\ \emph {et~al.}(2020)\citenamefont
  {Shahnazaryan}, \citenamefont {Kozin}, \citenamefont {Shelykh}, \citenamefont
  {Iorsh},\ and\ \citenamefont {Kyriienko}}]{Shahnazaryan:PRB102(2020)}%
  \BibitemOpen
  \bibfield  {author} {\bibinfo {author} {\bibfnamefont {V.}~\bibnamefont
  {Shahnazaryan}}, \bibinfo {author} {\bibfnamefont {V.~K.}\ \bibnamefont
  {Kozin}}, \bibinfo {author} {\bibfnamefont {I.~A.}\ \bibnamefont {Shelykh}},
  \bibinfo {author} {\bibfnamefont {I.~V.}\ \bibnamefont {Iorsh}},\ and\
  \bibinfo {author} {\bibfnamefont {O.}~\bibnamefont {Kyriienko}},\ }\href
  {https://doi.org/10.1103/physrevb.102.115310} {\bibfield  {journal} {\bibinfo
   {journal} {Phys. Rev. B}\ }\textbf {\bibinfo {volume} {102}},\ \bibinfo
  {pages} {115310} (\bibinfo {year} {2020})}\BibitemShut {NoStop}%
\bibitem [{\citenamefont {Shahnazaryan}\ \emph {et~al.}(2016)\citenamefont
  {Shahnazaryan}, \citenamefont {Shelykh},\ and\ \citenamefont
  {Kyriienko}}]{Shahnazaryan2016}%
  \BibitemOpen
  \bibfield  {author} {\bibinfo {author} {\bibfnamefont {V.}~\bibnamefont
  {Shahnazaryan}}, \bibinfo {author} {\bibfnamefont {I.~A.}\ \bibnamefont
  {Shelykh}},\ and\ \bibinfo {author} {\bibfnamefont {O.}~\bibnamefont
  {Kyriienko}},\ }\href {https://doi.org/10.1103/PhysRevB.93.245302} {\bibfield
   {journal} {\bibinfo  {journal} {Phys. Rev. B}\ }\textbf {\bibinfo {volume}
  {93}},\ \bibinfo {pages} {245302} (\bibinfo {year} {2016})}\BibitemShut
  {NoStop}%
\bibitem [{\citenamefont {Yagafarov}\ \emph {et~al.}(2020)\citenamefont
  {Yagafarov}, \citenamefont {Sannikov}, \citenamefont {Zasedatelev},
  \citenamefont {Georgiou}, \citenamefont {Baranikov}, \citenamefont
  {Kyriienko}, \citenamefont {Shelykh}, \citenamefont {Gai}, \citenamefont
  {Shen}, \citenamefont {Lidzey},\ and\ \citenamefont
  {Lagoudakis}}]{Yagafarov2020}%
  \BibitemOpen
  \bibfield  {author} {\bibinfo {author} {\bibfnamefont {T.}~\bibnamefont
  {Yagafarov}}, \bibinfo {author} {\bibfnamefont {D.}~\bibnamefont {Sannikov}},
  \bibinfo {author} {\bibfnamefont {A.}~\bibnamefont {Zasedatelev}}, \bibinfo
  {author} {\bibfnamefont {K.}~\bibnamefont {Georgiou}}, \bibinfo {author}
  {\bibfnamefont {A.}~\bibnamefont {Baranikov}}, \bibinfo {author}
  {\bibfnamefont {O.}~\bibnamefont {Kyriienko}}, \bibinfo {author}
  {\bibfnamefont {I.}~\bibnamefont {Shelykh}}, \bibinfo {author} {\bibfnamefont
  {L.}~\bibnamefont {Gai}}, \bibinfo {author} {\bibfnamefont {Z.}~\bibnamefont
  {Shen}}, \bibinfo {author} {\bibfnamefont {D.}~\bibnamefont {Lidzey}},\ and\
  \bibinfo {author} {\bibfnamefont {P.}~\bibnamefont {Lagoudakis}},\ }\href
  {https://doi.org/10.1038/s42005-019-0278-6} {\bibfield  {journal} {\bibinfo
  {journal} {Commun. Phys.}\ }\textbf {\bibinfo {volume} {3}},\ \bibinfo
  {pages} {18} (\bibinfo {year} {2020})}\BibitemShut {NoStop}%
\bibitem [{\citenamefont {Betzold}\ \emph {et~al.}(2020)\citenamefont
  {Betzold}, \citenamefont {Dusel}, \citenamefont {Kyriienko}, \citenamefont
  {Dietrich}, \citenamefont {Klembt}, \citenamefont {Ohmer}, \citenamefont
  {Fischer}, \citenamefont {Shelykh}, \citenamefont {Schneider},\ and\
  \citenamefont {H{\"o}fling}}]{Betzold2020}%
  \BibitemOpen
  \bibfield  {author} {\bibinfo {author} {\bibfnamefont {S.}~\bibnamefont
  {Betzold}}, \bibinfo {author} {\bibfnamefont {M.}~\bibnamefont {Dusel}},
  \bibinfo {author} {\bibfnamefont {O.}~\bibnamefont {Kyriienko}}, \bibinfo
  {author} {\bibfnamefont {C.~P.}\ \bibnamefont {Dietrich}}, \bibinfo {author}
  {\bibfnamefont {S.}~\bibnamefont {Klembt}}, \bibinfo {author} {\bibfnamefont
  {J.}~\bibnamefont {Ohmer}}, \bibinfo {author} {\bibfnamefont
  {U.}~\bibnamefont {Fischer}}, \bibinfo {author} {\bibfnamefont {I.~A.}\
  \bibnamefont {Shelykh}}, \bibinfo {author} {\bibfnamefont {C.}~\bibnamefont
  {Schneider}},\ and\ \bibinfo {author} {\bibfnamefont {S.}~\bibnamefont
  {H{\"o}fling}},\ }\href {https://doi.org/10.1021/acsphotonics.9b01300}
  {\bibfield  {journal} {\bibinfo  {journal} {ACS Photonics}\ }\textbf
  {\bibinfo {volume} {7}},\ \bibinfo {pages} {384} (\bibinfo {year}
  {2020})}\BibitemShut {NoStop}%
\bibitem [{\citenamefont {Wild}\ \emph {et~al.}(2018)\citenamefont {Wild},
  \citenamefont {Shahmoon}, \citenamefont {Yelin},\ and\ \citenamefont
  {Lukin}}]{Wild2018}%
  \BibitemOpen
  \bibfield  {author} {\bibinfo {author} {\bibfnamefont {D.~S.}\ \bibnamefont
  {Wild}}, \bibinfo {author} {\bibfnamefont {E.}~\bibnamefont {Shahmoon}},
  \bibinfo {author} {\bibfnamefont {S.~F.}\ \bibnamefont {Yelin}},\ and\
  \bibinfo {author} {\bibfnamefont {M.~D.}\ \bibnamefont {Lukin}},\ }\href
  {https://doi.org/10.1103/PhysRevLett.121.123606} {\bibfield  {journal}
  {\bibinfo  {journal} {Phys. Rev. Lett.}\ }\textbf {\bibinfo {volume} {121}},\
  \bibinfo {pages} {123606} (\bibinfo {year} {2018})}\BibitemShut {NoStop}%
\bibitem [{\citenamefont {Sich}\ \emph {et~al.}(2012)\citenamefont {Sich},
  \citenamefont {Krizhanovskii}, \citenamefont {Skolnick}, \citenamefont
  {Gorbach}, \citenamefont {Hartley}, \citenamefont {Skryabin}, \citenamefont
  {Cerda-M{\'e}ndez}, \citenamefont {Biermann}, \citenamefont {Hey},\ and\
  \citenamefont {Santos}}]{Sich2012}%
  \BibitemOpen
  \bibfield  {author} {\bibinfo {author} {\bibfnamefont {M.}~\bibnamefont
  {Sich}}, \bibinfo {author} {\bibfnamefont {D.~N.}\ \bibnamefont
  {Krizhanovskii}}, \bibinfo {author} {\bibfnamefont {M.~S.}\ \bibnamefont
  {Skolnick}}, \bibinfo {author} {\bibfnamefont {A.~V.}\ \bibnamefont
  {Gorbach}}, \bibinfo {author} {\bibfnamefont {R.}~\bibnamefont {Hartley}},
  \bibinfo {author} {\bibfnamefont {D.~V.}\ \bibnamefont {Skryabin}}, \bibinfo
  {author} {\bibfnamefont {E.~A.}\ \bibnamefont {Cerda-M{\'e}ndez}}, \bibinfo
  {author} {\bibfnamefont {K.}~\bibnamefont {Biermann}}, \bibinfo {author}
  {\bibfnamefont {R.}~\bibnamefont {Hey}},\ and\ \bibinfo {author}
  {\bibfnamefont {P.~V.}\ \bibnamefont {Santos}},\ }\href
  {https://doi.org/10.1038/nphoton.2011.267} {\bibfield  {journal} {\bibinfo
  {journal} {Nature Photonics}\ }\textbf {\bibinfo {volume} {6}},\ \bibinfo
  {pages} {50} (\bibinfo {year} {2012})}\BibitemShut {NoStop}%
\bibitem [{\citenamefont {Walker}\ \emph {et~al.}(2017)\citenamefont {Walker},
  \citenamefont {Tinkler}, \citenamefont {Royall}, \citenamefont {Skryabin},
  \citenamefont {Farrer}, \citenamefont {Ritchie}, \citenamefont {Skolnick},\
  and\ \citenamefont {Krizhanovskii}}]{Walker2017}%
  \BibitemOpen
  \bibfield  {author} {\bibinfo {author} {\bibfnamefont {P.~M.}\ \bibnamefont
  {Walker}}, \bibinfo {author} {\bibfnamefont {L.}~\bibnamefont {Tinkler}},
  \bibinfo {author} {\bibfnamefont {B.}~\bibnamefont {Royall}}, \bibinfo
  {author} {\bibfnamefont {D.~V.}\ \bibnamefont {Skryabin}}, \bibinfo {author}
  {\bibfnamefont {I.}~\bibnamefont {Farrer}}, \bibinfo {author} {\bibfnamefont
  {D.~A.}\ \bibnamefont {Ritchie}}, \bibinfo {author} {\bibfnamefont {M.~S.}\
  \bibnamefont {Skolnick}},\ and\ \bibinfo {author} {\bibfnamefont {D.~N.}\
  \bibnamefont {Krizhanovskii}},\ }\href
  {https://doi.org/10.1103/PhysRevLett.119.097403} {\bibfield  {journal}
  {\bibinfo  {journal} {Phys. Rev. Lett.}\ }\textbf {\bibinfo {volume} {119}},\
  \bibinfo {pages} {097403} (\bibinfo {year} {2017})}\BibitemShut {NoStop}%
\bibitem [{\citenamefont {Ma\^{\i}tre}\ \emph {et~al.}(2020)\citenamefont
  {Ma\^{\i}tre}, \citenamefont {Lerario}, \citenamefont {Medeiros},
  \citenamefont {Claude}, \citenamefont {Glorieux}, \citenamefont {Giacobino},
  \citenamefont {Pigeon},\ and\ \citenamefont {Bramati}}]{Maitre2020}%
  \BibitemOpen
  \bibfield  {author} {\bibinfo {author} {\bibfnamefont {A.}~\bibnamefont
  {Ma\^{\i}tre}}, \bibinfo {author} {\bibfnamefont {G.}~\bibnamefont
  {Lerario}}, \bibinfo {author} {\bibfnamefont {A.}~\bibnamefont {Medeiros}},
  \bibinfo {author} {\bibfnamefont {F.}~\bibnamefont {Claude}}, \bibinfo
  {author} {\bibfnamefont {Q.}~\bibnamefont {Glorieux}}, \bibinfo {author}
  {\bibfnamefont {E.}~\bibnamefont {Giacobino}}, \bibinfo {author}
  {\bibfnamefont {S.}~\bibnamefont {Pigeon}},\ and\ \bibinfo {author}
  {\bibfnamefont {A.}~\bibnamefont {Bramati}},\ }\href
  {https://doi.org/10.1103/PhysRevX.10.041028} {\bibfield  {journal} {\bibinfo
  {journal} {Phys. Rev. X}\ }\textbf {\bibinfo {volume} {10}},\ \bibinfo
  {pages} {041028} (\bibinfo {year} {2020})}\BibitemShut {NoStop}%
\bibitem [{\citenamefont {Bradley}\ \emph {et~al.}(1997)\citenamefont
  {Bradley}, \citenamefont {Sackett},\ and\ \citenamefont
  {Hulet}}]{Bradley1997}%
  \BibitemOpen
  \bibfield  {author} {\bibinfo {author} {\bibfnamefont {C.~C.}\ \bibnamefont
  {Bradley}}, \bibinfo {author} {\bibfnamefont {C.~A.}\ \bibnamefont
  {Sackett}},\ and\ \bibinfo {author} {\bibfnamefont {R.~G.}\ \bibnamefont
  {Hulet}},\ }\href {https://doi.org/10.1103/PhysRevLett.78.985} {\bibfield
  {journal} {\bibinfo  {journal} {Phys. Rev. Lett.}\ }\textbf {\bibinfo
  {volume} {78}},\ \bibinfo {pages} {985} (\bibinfo {year} {1997})}\BibitemShut
  {NoStop}%
\bibitem [{\citenamefont {Almand-Hunter}\ \emph {et~al.}(2014)\citenamefont
  {Almand-Hunter}, \citenamefont {Li}, \citenamefont {Cundiff}, \citenamefont
  {Mootz}, \citenamefont {Kira},\ and\ \citenamefont
  {Koch}}]{Almand-Hunter2014}%
  \BibitemOpen
  \bibfield  {author} {\bibinfo {author} {\bibfnamefont {A.~E.}\ \bibnamefont
  {Almand-Hunter}}, \bibinfo {author} {\bibfnamefont {H.}~\bibnamefont {Li}},
  \bibinfo {author} {\bibfnamefont {S.~T.}\ \bibnamefont {Cundiff}}, \bibinfo
  {author} {\bibfnamefont {M.}~\bibnamefont {Mootz}}, \bibinfo {author}
  {\bibfnamefont {M.}~\bibnamefont {Kira}},\ and\ \bibinfo {author}
  {\bibfnamefont {S.~W.}\ \bibnamefont {Koch}},\ }\href
  {https://doi.org/10.1038/nature12994} {\bibfield  {journal} {\bibinfo
  {journal} {Nature}\ }\textbf {\bibinfo {volume} {506}},\ \bibinfo {pages}
  {471} (\bibinfo {year} {2014})}\BibitemShut {NoStop}%
\bibitem [{\citenamefont {Khaykovich}\ \emph {et~al.}(2002)\citenamefont
  {Khaykovich}, \citenamefont {Schreck}, \citenamefont {Ferrari}, \citenamefont
  {Bourdel}, \citenamefont {Cubizolles}, \citenamefont {Carr}, \citenamefont
  {Castin},\ and\ \citenamefont {Salomon}}]{Khaykovich2002}%
  \BibitemOpen
  \bibfield  {author} {\bibinfo {author} {\bibfnamefont {L.}~\bibnamefont
  {Khaykovich}}, \bibinfo {author} {\bibfnamefont {F.}~\bibnamefont {Schreck}},
  \bibinfo {author} {\bibfnamefont {G.}~\bibnamefont {Ferrari}}, \bibinfo
  {author} {\bibfnamefont {T.}~\bibnamefont {Bourdel}}, \bibinfo {author}
  {\bibfnamefont {J.}~\bibnamefont {Cubizolles}}, \bibinfo {author}
  {\bibfnamefont {L.~D.}\ \bibnamefont {Carr}}, \bibinfo {author}
  {\bibfnamefont {Y.}~\bibnamefont {Castin}},\ and\ \bibinfo {author}
  {\bibfnamefont {C.}~\bibnamefont {Salomon}},\ }\href
  {https://doi.org/10.1126/science.1071021} {\bibfield  {journal} {\bibinfo
  {journal} {Science}\ }\textbf {\bibinfo {volume} {296}},\ \bibinfo {pages}
  {1290} (\bibinfo {year} {2002})},\ \Eprint
  {https://arxiv.org/abs/https://www.science.org/doi/pdf/10.1126/science.1071021}
  {https://www.science.org/doi/pdf/10.1126/science.1071021} \BibitemShut
  {NoStop}%
\bibitem [{\citenamefont {Firstenberg}\ \emph {et~al.}(2013)\citenamefont
  {Firstenberg}, \citenamefont {Peyronel}, \citenamefont {Liang}, \citenamefont
  {Gorshkov}, \citenamefont {Lukin},\ and\ \citenamefont
  {Vuleti{\'{c}}}}]{Firstenberg2013}%
  \BibitemOpen
  \bibfield  {author} {\bibinfo {author} {\bibfnamefont {O.}~\bibnamefont
  {Firstenberg}}, \bibinfo {author} {\bibfnamefont {T.}~\bibnamefont
  {Peyronel}}, \bibinfo {author} {\bibfnamefont {Q.-Y.}\ \bibnamefont {Liang}},
  \bibinfo {author} {\bibfnamefont {A.~V.}\ \bibnamefont {Gorshkov}}, \bibinfo
  {author} {\bibfnamefont {M.~D.}\ \bibnamefont {Lukin}},\ and\ \bibinfo
  {author} {\bibfnamefont {V.}~\bibnamefont {Vuleti{\'{c}}}},\ }\href
  {https://doi.org/10.1038/nature12512} {\bibfield  {journal} {\bibinfo
  {journal} {Nature}\ }\textbf {\bibinfo {volume} {502}},\ \bibinfo {pages}
  {71} (\bibinfo {year} {2013})}\BibitemShut {NoStop}%
\bibitem [{\citenamefont {Kyriienko}\ and\ \citenamefont
  {Liew}(2016)}]{KyriienkoLiew2016}%
  \BibitemOpen
  \bibfield  {author} {\bibinfo {author} {\bibfnamefont {O.}~\bibnamefont
  {Kyriienko}}\ and\ \bibinfo {author} {\bibfnamefont {T.~C.~H.}\ \bibnamefont
  {Liew}},\ }\href {https://doi.org/10.1103/PhysRevB.93.035301} {\bibfield
  {journal} {\bibinfo  {journal} {Phys. Rev. B}\ }\textbf {\bibinfo {volume}
  {93}},\ \bibinfo {pages} {035301} (\bibinfo {year} {2016})}\BibitemShut
  {NoStop}%
\bibitem [{\citenamefont {Ghosh}\ \emph {et~al.}(2021)\citenamefont {Ghosh},
  \citenamefont {Krisnanda}, \citenamefont {Paterek},\ and\ \citenamefont
  {Liew}}]{Ghosh2021}%
  \BibitemOpen
  \bibfield  {author} {\bibinfo {author} {\bibfnamefont {S.}~\bibnamefont
  {Ghosh}}, \bibinfo {author} {\bibfnamefont {T.}~\bibnamefont {Krisnanda}},
  \bibinfo {author} {\bibfnamefont {T.}~\bibnamefont {Paterek}},\ and\ \bibinfo
  {author} {\bibfnamefont {T.~C.~H.}\ \bibnamefont {Liew}},\ }\href
  {https://doi.org/10.1038/s42005-021-00606-3} {\bibfield  {journal} {\bibinfo
  {journal} {Commun. Phys.}\ }\textbf {\bibinfo {volume} {4}},\ \bibinfo
  {pages} {105} (\bibinfo {year} {2021})}\BibitemShut {NoStop}%
\bibitem [{\citenamefont {Xu}\ \emph {et~al.}(2021)\citenamefont {Xu},
  \citenamefont {Krisnanda}, \citenamefont {Verstraelen}, \citenamefont
  {Liew},\ and\ \citenamefont {Ghosh}}]{XuLiew2021}%
  \BibitemOpen
  \bibfield  {author} {\bibinfo {author} {\bibfnamefont {H.}~\bibnamefont
  {Xu}}, \bibinfo {author} {\bibfnamefont {T.}~\bibnamefont {Krisnanda}},
  \bibinfo {author} {\bibfnamefont {W.}~\bibnamefont {Verstraelen}}, \bibinfo
  {author} {\bibfnamefont {T.~C.~H.}\ \bibnamefont {Liew}},\ and\ \bibinfo
  {author} {\bibfnamefont {S.}~\bibnamefont {Ghosh}},\ }\href
  {https://doi.org/10.1103/PhysRevB.103.195302} {\bibfield  {journal} {\bibinfo
   {journal} {Phys. Rev. B}\ }\textbf {\bibinfo {volume} {103}},\ \bibinfo
  {pages} {195302} (\bibinfo {year} {2021})}\BibitemShut {NoStop}%
\bibitem [{\citenamefont {Kuriakose}\ \emph {et~al.}(2022)\citenamefont
  {Kuriakose}, \citenamefont {Walker}, \citenamefont {Dowling}, \citenamefont
  {Kyriienko}, \citenamefont {Shelykh}, \citenamefont {St-Jean}, \citenamefont
  {Zambon}, \citenamefont {Lema{\^i}tre}, \citenamefont {Sagnes}, \citenamefont
  {Legratiet}, \citenamefont {Harouri}, \citenamefont {Ravets}, \citenamefont
  {Skolnick}, \citenamefont {Amo}, \citenamefont {Bloch},\ and\ \citenamefont
  {Krizhanovskii}}]{Kuriakose2022}%
  \BibitemOpen
  \bibfield  {author} {\bibinfo {author} {\bibfnamefont {T.}~\bibnamefont
  {Kuriakose}}, \bibinfo {author} {\bibfnamefont {P.~M.}\ \bibnamefont
  {Walker}}, \bibinfo {author} {\bibfnamefont {T.}~\bibnamefont {Dowling}},
  \bibinfo {author} {\bibfnamefont {O.}~\bibnamefont {Kyriienko}}, \bibinfo
  {author} {\bibfnamefont {I.~A.}\ \bibnamefont {Shelykh}}, \bibinfo {author}
  {\bibfnamefont {P.}~\bibnamefont {St-Jean}}, \bibinfo {author} {\bibfnamefont
  {N.~C.}\ \bibnamefont {Zambon}}, \bibinfo {author} {\bibfnamefont
  {A.}~\bibnamefont {Lema{\^i}tre}}, \bibinfo {author} {\bibfnamefont
  {I.}~\bibnamefont {Sagnes}}, \bibinfo {author} {\bibfnamefont
  {L.}~\bibnamefont {Legratiet}}, \bibinfo {author} {\bibfnamefont
  {A.}~\bibnamefont {Harouri}}, \bibinfo {author} {\bibfnamefont
  {S.}~\bibnamefont {Ravets}}, \bibinfo {author} {\bibfnamefont {M.~S.}\
  \bibnamefont {Skolnick}}, \bibinfo {author} {\bibfnamefont {A.}~\bibnamefont
  {Amo}}, \bibinfo {author} {\bibfnamefont {J.}~\bibnamefont {Bloch}},\ and\
  \bibinfo {author} {\bibfnamefont {D.~N.}\ \bibnamefont {Krizhanovskii}},\
  }\bibfield  {journal} {\bibinfo  {journal} {Nature Photonics}\ }\href
  {https://doi.org/10.1038/s41566-022-01019-6} {10.1038/s41566-022-01019-6}
  (\bibinfo {year} {2022})\BibitemShut {NoStop}%
\end{thebibliography}%



\end{document}